\documentclass[conference, 9pt]{IEEEtran}
\IEEEoverridecommandlockouts
\usepackage{cite}
\def\BibTeX{{\rm B\kern-.05em{\sc i\kern-.025em b}\kern-.08em
    T\kern-.1667em\lower.7ex\hbox{E}\kern-.125emX}}

\usepackage{graphicx,dsfont,amssymb,booktabs} % For formal tables

\usepackage[noend]{algpseudocode}
\usepackage{algorithm}
\usepackage{amsmath}
\usepackage{diagbox}
\usepackage{subcaption} 

\usepackage{color, colortbl}
\usepackage{tabulary}
\usepackage{tabularx}
\usepackage{makecell}
\usepackage[textsize=scriptsize]{todonotes}
\usepackage{multirow}
\usepackage{ctable} % for \specialrule command
\usepackage{hhline}

\usepackage{pgfplots}
\usepackage{tikz}
\usetikzlibrary{
	arrows.meta,
	colorbrewer,
	decorations.pathreplacing,
	automata, 
	positioning }

\usepackage{listings}
\usepackage{pgfmath}

\usepackage{balance}

\usepackage{xcolor}

\newcommand{\gSPICE}{gSPICE}

\newcommand{\gSPICESH}{gSPICE-H}
\newcommand{\gSPICESM}{gSPICE-M}
\newcommand{\gSPICEST}{gSPICE-T}
\newcommand{\gSPICESF}{gSPICE-F}

\begin{document}

\title{\gSPICE: Model-Based Event Shedding in Complex Event Processing}

\vspace{-0.4cm}
\author{\IEEEauthorblockN{ Ahmad Slo}
\IEEEauthorblockA{\textit{University of Stuttgart} \\
Stuttgart, Germany \\
ahmad.slo@ipvs.uni-stuttgart.de}
\and
\IEEEauthorblockN{ Sukanya Bhowmik}
\IEEEauthorblockA{\textit{University of Potsdam} \\
Potsdam, Germany \\
sukanya.bhowmik@uni-potsdam.de}
\and
\IEEEauthorblockN{Kurt Rothermel}
	\IEEEauthorblockA{\textit{University of Stuttgart} \\
		Stuttgart, Germany \\
		kurt.rothermel@ipvs.uni-stuttgart.de}}

\maketitle
	\vspace{-0.4cm}
\begin{abstract}

Overload situations, in the presence of resource limitations, in complex event processing (CEP) systems are typically handled using load shedding to maintain a given latency bound. However, load shedding might negatively impact the quality of results (QoR). To minimize the shedding impact on QoR, CEP researchers propose shedding approaches that drop events/internal state with the lowest importances/utilities. In both black-box and white-box shedding approaches, different features are used to predict these utilities. In this work, we propose a novel black-box shedding approach that uses a new set of features to drop events from the input event stream  to maintain a given latency bound. Our approach uses a probabilistic model to predict these event utilities. Moreover, our approach uses Zobrist hashing and well-known machine learning models, e.g., decision trees and random forests,  to handle the predicted event utilities.
Through extensive evaluations on several synthetic and two real-world datasets and a representative set of CEP queries, we show that, in the majority of cases, our load shedding approach outperforms state-of-the-art black-box load shedding approaches, w.r.t. QoR.

\end{abstract}


\section{Introduction}

Complex event processing (CEP) is a powerful paradigm to derive high level information (called complex events) from streams of primitive events. CEP systems are used in many applications, e.g., stock trading, transportation, network monitoring, retail management \cite{espice, Olston:2003:AFC:872757.872825, Wu:2006:HCE:1142473.1142520}.  CEP operators correlate events, in accordance with defined CEP patterns, in the input event stream to detect complex events. The detected complex events may represent important situations for the application. In many applications, complex events must be detected within a certain latency bound, rendering them useless otherwise~\cite{10.1145/3361525.3361551, Quoc:2017:SAC:3135974.3135989} . Therefore,  a CEP system must be able to maintain a given latency bound while correlating events.

In overload cases, a CEP operator receives more events than it can process. Therefore, many researchers in the CEP domain propose to shed load by dropping either events or a portion of the operator's internal state to maintain a given latency bound \cite{He2014OnLS, espice, pspice, bo:2020}. Load shedding may be used in CEP applications that tolerate imprecise complex event detections, for example, network monitoring \cite{Olston:2003:AFC:872757.872825},  computing cluster monitoring \cite{bo:2020}, soccer analysis, and stock trading \cite{espice, pspice}. In  \cite{pspice, bo:2020}, the authors propose white-box load shedding approaches, where the load shedder has access to the operator's internal state. While in \cite{He2014OnLS, espice}, the authors propose black-box load shedding approaches, where the operator's internal state is not revealed. A clear advantage of a black-box shedding approach is that the shedding functionality can be easily added to a CEP operator with minimal overhead on a domain expert. In fact, there is no need to modify the operator implementation. As a result, such a load shedder, that performs shedding agnostic to the operator implementation, has a universal appeal. Therefore, in this work, we focus on a black-box shedding approach that performs load shedding based on input event streams. 

Of course, shedding load may adversely impact the quality of results (QoR). Therefore, an  efficient load shedding approach must maintain the given latency bound while minimizing the negative shedding impact on QoR.

To predict the utility of events, the works in \cite{He2014OnLS, espice} use a limited set of features,  such as the event type, event order, and event frequency in patterns and input event stream. However, other important features may exist that help in accurately predicting the utility of an event. As a result, in this paper, we propose to explore a new set of features that go beyond the aforementioned ones.
We, first, define the \textit{predecessor pane} of event $e$ as the sequence of events that occur before an event $e$ in the input event stream.
We consider the predecessor pane an important feature since it indicates the current progress of pattern matching within the operator. Second, we consider the \textit{event content} an important feature since, in CEP, events in patterns are correlated while fulfilling certain predicates on the event content.
 To summarize, in this work, we propose a novel \textit{black-box} load shedding approach for CEP systems, called \gSPICE{}. In overload cases, \gSPICE{} drops \textit{events} from the input event stream of a CEP operator to maintain a given latency bound. To minimize the shedding impact on QoR, \gSPICE{} drops events with the lowest utilities, where it uses a probabilistic model to predict event utilities.
The model depends on three features---\textit{predecessor pane, event content, and event type}.

Using complex features such as the predecessor pane and event content to predict the event utility, on the one hand, might improve the prediction accuracy. But on the other hand, it might result in a heavyweight model that consumes high computational time to predict event utilities.
To design an effective load shedding approach in CEP, the approach must have the following two properties. 1) The approach must accurately predict the utility of events as failing to do so may result in dropping important events, hence negatively impacting QoR. 2) The approach should shed load with low computational overhead, i.e., should be a lightweight load shedding approach. The reason behind this is that the computing resources assigned to a CEP operator to match patterns are also used to take the shedding decision. Thus, a high load shedding overhead results in wasting more computing resources designated to match patterns. This increases the need to drop more events which may negatively impact QoR.

As a result, our contributions in this paper are as follows:
\begin{itemize}
	\item We propose \gSPICE{}, a black-box load shedding approach for CEP systems, that drops events from the input event stream of an overloaded CEP operator to maintain a given latency bound. \gSPICE{} uses a probabilistic model to predict event utilities depending on the following features: 1) event type, 2) predecessor pane, and 3) event content.
	
	\item  We develop a data structure that depends on the Zobrist hashing \cite{Zobrist1990ANH} to efficiently store the event utilities. This data structure enables \gSPICE{} to perform load shedding in a lightweight manner.
	
	\item  We also propose to use well-known machine learning approaches, e.g., decision trees or random forests, to estimate event utilities.
	
	\item  We perform extensive evaluations on several real world and synthetic datasets and a representative set of queries to show the performance of \gSPICE, w.r.t. its impact on QoR, and compare
	its performance with state-of-the-art black-box shedding approaches. 
	 
\end{itemize}

\section{Preliminaries and Problem Statement}
\label{sec:background}
\subsection{Complex Event Processing}
A complex event processing (CEP) system is represented as a directed acyclic graph (DAG) of CEP operators. Operators correlate primitive events in the input event stream  (denoted by $S_{in}$)  to detect patterns (i.e., complex events).  A primitive event (or simply, event) is the basic data element in CEP systems that represents the occurrence of an application-related situation. 
%A primitive event (or simply, event) is atomic (i.e., happens completely or not at all) and happens at a certain point in time \cite{ch1994snoop, Zimmer:1999:SCE:846218.847253}. 
An event consists of meta-data and attribute-value pairs. The meta-data contains the event type and timestamp, while the attribute-value pairs represent the actual event data (i.e., the event content). For example, the type (denoted by $T_e$) of an event $e$ might represent a company name in a stock application or a player ID in a soccer application. We refer to the set of all event types in the input event stream as $\mathbb{T}$. 
%Event timestamp represents the point in time when the event occurred. 
An attribute $E_e$ of an event $e$ might, for example, represent the stock quote of a company or a player position in the above two applications.
 We refer to the set of all event attributes as $\mathbb{E}_e$.

In this work, we focus on a CEP system consisting of a single operator that matches multiple patterns (i.e., multiple queries) $\mathbb{Q}= \{q_i: 1\le i \le n \}$, where $n$ is the number of patterns.  Since patterns might have different importances, each pattern has a weight (given by a domain expert) reflecting its importance as follows: $ \mathbb{W}_\mathbb{Q} = \{ w_{q_i} : 1 \le i \le n\}$, where $w_{q_i}$ is the weight of pattern $q_i$. In CEP, to detect complex events, it is common to correlate (a.k.a. process) together only events that occur within a certain interval (a.k.a. window) \cite{Balkesen:2013:RRI:2488222.2488257}. Processing events within windows results in producing partial matches where a partial match (short PM) represents a matched part of  a pattern. A PM might complete and become a complex event. When a window closes, all PMs produced within the window are abandoned. To clarify the system model, let us introduce the following example.

\textbf{Example 1.}
In a retail management application, every item  is equipped with an  RFID tag where there are three primitive events that may be generated for each item by RFID readers \cite{Wu:2006:HCE:1142473.1142520}: 1) a shelf reading event ($R$) if an item is removed from a shelf, 2) a counter reading event ($C$) if the item is checked out on the counter, 3) an exit reading event ($X$) if the item is carried outside the retail store.   To detect shoplifting, a CEP operator matches a pattern $q$ which correlates events generated by RFID readers. Pattern $q$ is defined as follows: generate a complex event if there exists a shelf reading event $R$  and an exit reading event $X$ for an item $M$ but there is no counter reading event $C$ for the item $M$  within a certain time, e.g., two hours (i.e., window size $ws= 2$ hours).    
We may write this pattern as a sequence event operator with the negation event operator \cite{ch1994snoop, Wu:2006:HCE:1142473.1142520}:
\begin{equation*}[q]
\begin{aligned} 
&\textbf{pattern} \quad \mathbf{seq}~ (R; !C; X) \\
&\qquad \mathbf{where}~  R.ID= C.ID ~ \mathbf{and} ~ R.ID= X.ID  \\
&\qquad \textbf{within}~ 2 ~\text{hours}  
\end{aligned}
\label{eq:example1}
%\vspace{-0.3cm}
\end{equation*}

In this example, the set of patterns $\mathbb{Q}= \{q\}$ and the event types represent the shelf reading $R$, counter reading $C$, and exit reading $X$, hence $\mathbb{T}= \{R, C, X\}$. Moreover, in this example, there is only one event attribute which is the item ID, hence the set of event attributes $\mathbb{E}_e= \{ ID \}$.  Let us assume an input event stream $S_{in}$ with the following events, as depicted in Figure \ref{fig:event-context}: $ X_4 ~ R_3 ~ X_2 ~ C_1 ~  R_0$, where event $E_i$ is of type $E$ at position $i$ in the input event stream, i.e., $i$ defines the event order in the input event stream $S_{in}$. We refer to $E_i$ as an instance of event $E$, e.g., $R_0$ is an instance of event $R$. Assume that a window $w$ contains the events $ X_4 ~ R_3 ~ X_2 ~ C_1 ~  R_0$. Processing events in $w$ to detect pattern $q$ is performed as follows. Processing event $R_0$ opens a PM $\gamma_1$ which is abandoned when processing event $C_1$ in $w$ since the event type $C$ is a negated event in pattern $q$, assuming that both events $R_0$ and $C_1$ are generated by the RFID readers for the same item (i.e., $R_0.ID= C_1.ID$). As $R_0$ and $C_1$ have updated the progress of PM $\gamma_1$, we refer to $R_0$ and $C_1$ as  the events that \textit{contribute} to PM $\gamma_1$. Processing Event $X_2$  in window $w$ does not result in any match since there exists no open PMs, hence event $X_2$ does not contribute to any PM. The event $R_3$ opens a new PM $\gamma_2$ which completes and becomes a complex event when processing event $X_4$ in window $w$, assuming $R_3.ID= X_4.ID$. Hence, processing window $w$ results in detecting  only one complex event $cplx_{34}= (R_3, X_4)$. We refer to $R_3$ and $X_4$ as events that \textit{contribute} to complex event  $cplx_{34}$. Moreover, we refer to $cplx_{34}$ as the complex event of pattern $q$, denoted by $cplx_{34} \subset q$. Since a pattern $q_i$ has a weight $w_{q_i}$, a complex event of pattern $q_i$ also has the weight $w_{q_i}$.

In overload cases, as mentioned above, we must drop events to maintain a given latency bound. Dropping events might, of course, degrade QoR which is represented by the number of false negatives and positives. A false negative is a missed complex event that should have been detected. While a false positive is a falsely detected complex event that should not be detected.  In Example 1, due to overload, assume that one event must be dropped from the input event stream $S_{in}$ to prevent violating a given latency bound. If we drop event $X_2$, it does not result in false negatives or positives. However, if event $R_3$ is dropped, the operator does not open PM $\gamma_2$. As a result, the operator cannot detect complex event $cplx_{34}$, which results in one false negative.  However,  if event $C_1$ is dropped, PM $\gamma_1$ is not abandoned. Hence, the operator detects a new complex event $cplx_{02}= (R_0, X_2)$, assuming $R_0.ID= X_2.ID$, which results in one false positive.

\subsection{Predecessor Pane}
We, now, define the predecessor pane. The predecessor pane (denoted by $\omega^e$) of an event $e$ represents a sequence of a certain number of events that occur before event $e$ in the input event stream $S_{in}$. The number of events in the predecessor pane is determined by the length of the predecessor pane (denoted by $L_\omega$).  The length of the predecessor pane  might be either time-based  or count-based. For example, a pane of length 5 seconds ($L_\omega= 5 ~seconds$), i.e., the predecessor pane $\omega^e$ of an event $e$ contains all events within the last 5 seconds from event $e$. A pane of length 100 events ($L_\omega= 100 ~events$), i.e., the predecessor pane $\omega^e$ of an event $e$ contains the last 100 events that occurred before event $e$ in the input event stream $S_{in}$. Without loss of generality, to simplify the presentation, next, we will assume that the predecessor pane length $L_\omega$ is count-based if not otherwise stated. 
The predecessor pane $\omega^{e_j}$ of event $e_j$ is formally defined as follows: $\omega^{e_j}= ( e_i: e_i \in S_{in}  ~\&~ j - L_\omega \le i < j )$ where $i$ and $j$ represent the event order in the input event stream $S_{in}$.

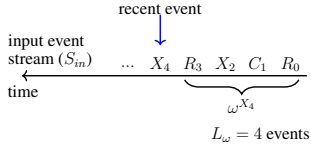
\begin{figure}[t]
	\centering
	\resizebox{0.50\linewidth}{!}{% !TeX root = ../../paper.tex

\definecolor{w1Color}{rgb}{1,0,0}
\definecolor{w2Color}{rgb}{137,0,255}
\definecolor{w3Color}{rgb}{0,0,1}
\tikzset{
	label style/.style={font=\huge\bf, color=black},
	rectangle style/.style ={color= blue, line width=1pt }, 
	arrow style right/.style={color=black, line width=1pt, -{>[scale=2.5, length=4, width=2]}}, 
	arrow style left/.style= {color=black, line width=1pt, {<[scale=2.5, length=4, width=2]}-}, 
%	line style/.style={ line width=1pt},
	node pattern style/.style= {shorten >=1pt, node distance=3cm,on grid,auto, line width=1pt,  initial distance= 0.5cm},	
	node pm style/.style= { shorten >=1pt, node distance=2 cm,on grid,auto, line width=1pt},	
	big font/.style={font=\fontsize{12.0}{13.0}\selectfont},
	every node/.style={big font},
	every state/.style={minimum size=0pt, inner sep=2pt}
}

\begin{tikzpicture}[xscale= 1,yscale=1]

%event stream
\newcommand{\xStart}{0}
\newcommand{\xEnd}{7}
\newcommand{\yStart}{0}
\newcommand{\xEventShift}{-0.8}

\path[big font] (\xEnd -0.4, \yStart) node[above] {$R_{0}$}  ++ (\xEventShift, 0) node[above] {$C_{1}$} ++ (\xEventShift, 0) node[above] {$X_{2}$} ++ (\xEventShift, 0) node[above] {$R_{3}$} ++ (\xEventShift, 0) node[above] {$X_{4}$}++ (\xEventShift, 0) node[above= 3pt] {$...$};

\draw [->, color=blue, line width=1pt, big font] (\xEnd + \xEventShift*4.5, \yStart + 1.5) node[above= -3pt, align=center,  color=black, big font] {recent event} -- (\xEnd + \xEventShift*5 + 0.4, \yStart + 0.75);
\draw [<-, line width=1pt,  big font] (\xStart, \yStart) -- (\xEnd, \yStart)  node [pos= 0.1, above= 2.5pt , align=left, name=txt] {input event \\stream ($S_{in}$)};
\node[below= 1.0cm of txt.west,anchor=west, big font] {time};

\draw[label style, line width=1pt,decorate,decoration={brace,amplitude=10pt},  big font] (\xEnd - 0.2,-0.1) -- (\xEnd + \xEventShift*4 + 0.2,-0.1) node[midway, below=10pt, name=context] {$\omega^{X_4}$};

\node[below left= 0.7cm  and 0.4cm of context.west,anchor=west] {$L_\omega= 4 $ events };

\end{tikzpicture}}
%	\vspace{-0.4cm}
	\caption{Predecessor pane.}
	\label{fig:event-context}
	\vspace{-0.6cm}
\end{figure}

Figure \ref{fig:event-context} depicts an example of input event stream $S_{in}$ and the  predecessor pane of event $X_4$. The pane is of length four, i.e., $L_\omega= 4$ events.  In the figure, the predecessor pane $\omega^{X_4}$ contains events $R_0$, $C_1$, $X_2$, and $R_3$, i.e., $\omega^{X_4}= (R_0, C_1, X_2, R_3)$. 

\textbf{Type Frequency in Predecessor Pane. }
Next, we define the type frequency in the predecessor pane. Type Frequency (denoted by $F$) in the predecessor pane $\omega^e$  of an event $e$ is a sequence representing the number of occurrences of each event type in the  predecessor pane $\omega^e$.  More formally, for an event $e$, the type frequency $F$ in the predecessor pane $\omega^e$ is defined as follows:
$$
F=( F_{T_i} : \forall~ T_i \in \mathbb{T},  F_{T_i}= \textstyle \sum_{\substack{e^\prime = \omega^e_j : ~ 0\le j < L_\omega\\ 
T_i= T_{e^\prime}} }^{} {1}) .
$$
In Figure \ref{fig:event-context}, in the predecessor pane $\omega^{X_4}$, there are two events of event type $R$, one event of type $C$, and one event of type $X$. Therefore, the type frequency in the predecessor pane $\omega^{X_4}$ is defined as follows: $F = (F_R, F_C, F_X)=  (2, 1, 1)$.

\gSPICE{} is a black-box load shedding approach where the CEP operator only reveals the detected complex events. Additionally, \gSPICE{} has access to events in the input event streams $S_{in}$. 
There exist several event operators and pattern matching semantics (a.k.a selection and consumption policies) in CEP systems \cite{ch1994snoop, Wu:2006:HCE:1142473.1142520, Zimmer:1999:SCE:846218.847253, cu2010tesla}. \gSPICE{} is a generic load shedding approach where we do not assume a specific event operator or selection and consumption policy.

\subsection{Problem Statement} 
An overloaded CEP operator must shed events to maintain a given latency bound, thus negatively impacting QoR. Therefore, the objective is to shed events in a way that has a minimum negative impact on QoR. 

Typically, an operator detects a set of patterns (i.e., $\mathbb{Q}$) with given weights (i.e., $\mathbb{W_Q}$). Therefore, for pattern $q_i \in \mathbb{Q}$, we define the number of false positives as $FP_{q_i}$ and the number of false negatives as $FN_{q_i}$.
Our  objective is to minimize the adverse impact on QoR, i.e., minimize the sum of the total number of weighted false positives and negatives, while dropping events such that the given latency bound $LB$ is met.  More formally, the objective is defined as follows.
\begin{equation}
\begin{aligned}
 minimize \quad  & \textstyle \sum_{q_i \in \mathbb{Q}} w_{q_i} * FP_{q_i} + \sum_{q_i \in \mathbb{Q}} w_{q_i} * FN_{q_i}	\\
\textrm{s.t.} \quad & \ l_e \le LB \quad \forall~ e \in S_{in},
\end{aligned}
\end{equation}
where $l_e$ is the latency of event $e$  that represents the sum of the queuing latency of $e$ and the time needed to process $e$.

\section{\gSPICE}
\label{sec:ls}
In this section, we present our proposed load shedding approach called \gSPICE. 
%\gSPICE{} drops events from the input event stream $S_{in}$ of a CEP operator. 
%Figure \ref{fig:ls-architecture} depicts the architecture of \gSPICE.
To perform load shedding in CEP, a load shedding strategy must perform three main tasks: 1) deciding when to drop events, 2) computing the percentage of events  to drop  (denoted by $\rho$) and the interval in which events must be dropped, and 3) determining which events to drop. 
%In Figure \ref{fig:ls-architecture}, the overload detector performs tasks 1 and 2. In overload cases, to avoid violating a given latency bound (LB), the overload detector commands the load shedder (LS) to drop $\rho\%$ of events from the input event stream every drop interval. 
In overload cases, to avoid violating a given latency bound (LB), the load shedder must drop $\rho\%$ of events from the input event stream every drop interval.  For example, the used window size might be used as a drop interval. 
Tasks 1 and 2 have been extensively researched in literature \cite{espice, Tatbul:2003:LSD:1315451.1315479}. Therefore, in this work, we focus on task 3, i.e., determining which events to drop.

To minimize the negative impact of event dropping  on QoR, \gSPICE{} must drop those events that have the lowest importance where the event importance is derived from the number of complex events to which the event contributes. If event $e$ contributes to a high number of complex events, event $e$ has high importance.
% However, if event $e$ does not contribute to any complex event, event $e$ has low importance. 
We refer to the event importance as the event utility. Higher is the event importance, higher is its utility and vice versa.

\gSPICE{} performs two main tasks: 1) model building and 2) load shedding. In the model building task, \gSPICE{} predicts event utilities based on selected features. During load shedding, \gSPICE{} drops events with the lowest utilities where it uses the event utilities predicted by the  model. Please recall, the utility of an event $e$ depends on the number of complex events to which event $e$ contributes.  The number of complex events to which event $e$ contributes is only known after processing event $e$. However, to decide which events to drop, we must identify event utilities before processing them and drop those with the lowest utilities.
%, hence reducing the negative impact of dropping on QoR and preventing latency bound violation. 
In fact, we need to predict the utility of each event $e$ before processing it at the operator.
Since, the shedder does not know the number of complex events to which an event will eventually contribute, we must depend on other features to assign utility to events.
Therefore, we first determine the features that are important to predict event utility in \gSPICE. Then, we explain the way \gSPICE{} predicts the event utilities.  Finally, we show how load shedding in \gSPICE{} is performed.

\subsection{Event Utility}
\label{sec:event-utility}

\gSPICE{} depends on three features to predict utility (denoted by $U_e$) of an event $e$ in the input event stream $S_{in}$: 1) event type $T_e \in \mathbb{T}$, 2) type frequency $F$ in the predecessor pane $\omega^e$, and 3) event attributes $\mathbb{E}_e$ (i.e., the event content).
Event attributes are important features for predicting the event utility since a CEP pattern usually correlates events that contain attributes that fulfill specific conditions. In Example 1, the pattern $q$ matches events with the same $ID$ (i.e., $ID$ is an event attribute). Therefore, event attributes might have a considerable influence on the event utility.
We consider only attributes with numerical values since more complex attributes (e.g., text or images) might considerably increase the load shedding overhead, adversely impacting QoR.
Event type $T_e$ and type frequency $F$ together represent important features to predict the event utility $U_e$ as well. 
Type frequency $F$ determines the importance of event $e$ of type $T_e$ since the type frequency (derived from the predecessor pane) contains information on events that happen before event $e$ in the input event stream $S_{in}$. Thus, it gives an indication of the number of open PMs and the current progress of these PMs (i.e., states of these PMs). This in turn gives an indication of the number of PMs to which event $e$ of type $T_e$ may contribute, hence the number of complex events to which event $e$ might contribute.  

\begin{figure}[t]
	\centering
	\resizebox{0.40\linewidth}{!}{% !TeX root = ../../paper.tex

\definecolor{w1Color}{rgb}{1,0,0}
\definecolor{w2Color}{rgb}{137,0,255}
\definecolor{w3Color}{rgb}{0,0,1}
\tikzset{
	label style/.style={font=\huge\bf, color=black},
	rectangle style/.style ={color= black, line width=1pt }, 
	arrow style right/.style={color=black, line width=1pt, -{>[scale=2.5, length=4, width=2]}}, 
	arrow style left/.style= {color=black, line width=1pt, {<[scale=2.5, length=4, width=2]}-}, 
%	line style/.style={ line width=1pt},
	node pattern style/.style= {shorten >=1pt, node distance=3cm,on grid,auto, line width=1pt,  initial distance= 0.5cm},	
	node pm style/.style= { shorten >=1pt, node distance=2 cm,on grid,auto, line width=1pt},	
	big font/.style={font=\fontsize{8.0}{9.0}\selectfont},
	every node/.style={big font},
	every state/.style={minimum size=0pt, inner sep=2pt}
}

\begin{tikzpicture}[xscale= 1,yscale=0.5]

\def\events{{"$X_4$","$R_3$", "$R_2$", "$R_1$", "$R_0$"}}
\def\XStart{0}
\def\YStart{0}
\def\SquareSize{1}
\foreach \n in {0,...,4}
{
	\ifthenelse{\equal{\n}{0}}
	{
		\draw[rectangle style, rounded corners=3pt, big font] (\n + \XStart, \YStart) rectangle (\n + \XStart + \SquareSize, \YStart+ \SquareSize) node [color=blue, pos= 0.5, big font] { \pgfmathparse{\events[\n]}\pgfmathresult};
	}
	{
		\draw[rectangle style, rounded corners=3pt, big font ] (\n + \XStart, \YStart) rectangle (\n + \XStart + \SquareSize, \YStart+ \SquareSize) node [color=black, pos= 0.5] {\pgfmathparse{\events[\n]}\pgfmathresult};
	};
};
\draw[label style, line width=1pt,decorate,decoration={brace,amplitude=10pt, mirror}, big font] (1,-0.1) -- (5,-0.1) node[midway, below=10pt, name=context] {$\omega^{X_4}$};
\node[blue] at (2.3, - 1.5  + \YStart ) {$\mathbf{(a)}$ };

\def\events{{"$X_4$","$C_3$", "$R_2$", "$R_1$", "$R_0$"}}
\def\YStart{-3.5}
\foreach \n in {0,...,4}
{
	\ifthenelse{\equal{\n}{0}}
	{
		\draw[rectangle style, rounded corners=3pt, big font] (\n + \XStart, \YStart) rectangle (\n + \XStart + \SquareSize, \YStart+ \SquareSize) node [color=blue, ,pos= 0.5] { \pgfmathparse{\events[\n]}\pgfmathresult};
	}
	{
		\draw[rectangle style, rounded corners=3pt,big font] (\n + \XStart, \YStart) rectangle (\n + \XStart + \SquareSize, \YStart+ \SquareSize) node [color=black, pos= 0.5] {\pgfmathparse{\events[\n]}\pgfmathresult};
	};
};
\draw[label style, line width=1pt,decorate,decoration={brace,amplitude=10pt, mirror}, big font] (1,-0.1 + \YStart) -- (5,-0.1 -0.1 + \YStart) node[midway, below=10pt, name=context] {$\omega^{X_4}$};
\node[blue] at (2.3, - 1.5  + \YStart ) {$\mathbf{(b)}$};

%\def\events{{"$X_4$","$C_3$", "$C_2$", "$R_1$", "$R_0$"}}
%\def\YStart{-5}
%\foreach \n in {0,...,4}
%{
%	\ifthenelse{\equal{\n}{0}}
%	{
%		\draw[rectangle style, rounded corners=5pt] (\n + \XStart, \YStart) rectangle (\n + \XStart + \SquareSize, \YStart+ \SquareSize) node [color=blue, pos= 0.5] { \pgfmathparse{\events[\n]}\pgfmathresult};
%	}
%	{
%		\draw[rectangle style, rounded corners=5pt ] (\n + \XStart, \YStart) rectangle (\n + \XStart + \SquareSize, \YStart+ \SquareSize) node [color=black, pos= 0.5] {\pgfmathparse{\events[\n]}\pgfmathresult};
%	};
%};
%\draw[label style, line width=1pt,decorate,decoration={brace,amplitude=10pt, mirror}] (1,-0.1 + \YStart) -- (5,-0.1 -0.1 + \YStart) node[midway, below=10pt, name=context] {$\mathbb{C}_{X_4}$};
%\node[blue] at (2.5, - 1  + \YStart ) {$\mathbf{(c)}$ };

\node[below= 0.5cm of context.west,anchor=west] {$L_\omega= 4 $ events };

\end{tikzpicture}}
%	\vspace{-0.4cm}
	\caption{Importance of the predecessor pane.}
	\label{fig:event-context-importance}
%	\vspace{-0.6cm}
\end{figure}
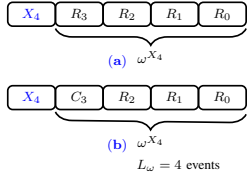

For example, using the shoplifting query in Example 1 (cf. Section \ref{sec:background}), Figure \ref{fig:event-context-importance} depicts examples of two different predecessor panes of length $L_\omega= 4$ events for the event $X_4$. In this example, let us assume that there are at most four events before the event $X_4$ in windows that contain event $X_4$, i.e., event $X_4$ might contribute only to PMs that are opened by the latest four events before event $X_4$. In Figure \ref{fig:event-context-importance}(a), the predecessor pane of event $X_4$ contains four events of type $R$ (i.e., $R_0$, $R_1$, $R_2$, $R_3$) where each event of type $R$ opens a new PM. Since there are no events of type $C$ before event $X_4$, event $X_4$ has a high probability to contribute to an open PM and results in detecting a complex event. Therefore, event $X_4$ should have a high utility in this example, i.e., Figure \ref{fig:event-context-importance}(a). 
In Figure \ref{fig:event-context-importance}(b), there are three events of type $R$ (i.e., $R_0$, $R_1$, $R_2$) and one event of type $C$ (i.e., $C_3$) in the predecessor pane  of event $X_4$. In this figure, event $C_3$ might abandon an already open PM. 
As a result, if events $C_3$  and $X_4$ are generated for the same item $M$ (i.e., $C_3.ID = X_4.ID$), event $X_4$ will not contribute to any PM since the PM that event $X_4$ might contribute to is abandoned by event $C_3$. Hence, in this case, event $X_4$ is not important, and its utility should be low.
However, if events $C_3$  and $X_4$ are generated for different items (i.e., $C_3.ID \ne X_4.ID$), event $X_4$ will  contribute to an open PM and results in detecting a complex event. Hence, $X_4$ should be assigned a high utility value.  As a result, in the above example, if the number of events of type $C$ increases, the utility of event $X_4$ might decrease.

As a result, we write the utility $U_e$ of event $e$ as a function (called utility function) of these three features (i.e., event type $T_e \in \mathbb{T}$, type frequency $F$ in the predecessor pane $\omega^e$, and event attributes $\mathbb{E}_e$) as shown in Equation \ref{eq:u_e}:
\begin{equation} 
\textstyle U_e= f(T_e, F, \mathbb{E}_e)
\label{eq:u_e}
\end{equation}

\subsection{Predicting Event Utility}
\label{sec:predicting_event_utility}
Now, we explain how to predict the event utility $U_e$, i.e., build the utility function shown in Equation \ref{eq:u_e}.  
\gSPICE{} predicts the event utility depending on gathered statistics. Therefore, next, we first show how \gSPICE{} gathers statistics and uses them to predict event utilities. Then, we explain the way \gSPICE{} handles the predicted utilities.
Let us first introduce the following simple examples.

\textbf{\textit{Example 2.}}
A CEP operator matches the pattern $q= seq (A;B)$. Assume that the input event stream contains only two event types $A$ and $B$ and events have only a single attribute $E_e$, i.e., $\mathbb{E}_e = \{E_e\}$. Moreover, assume that a predecessor pane of length 3 events (i.e., $L_\omega= 3$) is used.

\subsubsection{Gathering Statistics}
To predict the value of the utility function $f$, \gSPICE{} gathers statistics from the already processed events in the input event stream $S_p$. For each event $e \in S_p$,  \gSPICE{}  builds an observation on the set of complex events (denoted by $\mathbb{M}_e$) to which event $e$ contributes. The observation (denoted by $ob_e$) is of the following form: $ob_e \langle T_e, F, \mathbb{E}_e, \mathbb{M}_e \rangle$, $T_e$ is the type of event e, $F$  represents the type frequency in the predecessor pane $\omega^e$, and $\mathbb{E}_e$ is the set of attributes of event $e$. The observations $ob_e$ are stored in a set called the observation set $\mathbb{S}_e$, i.e., $ob_e \in \mathbb{S}_e$.
In Figure \ref{fig:observations-example}, Table \ref{tab:obsevations} shows observations gathered for pattern $q$ in Example 2. To simplify the presentation, the table shows observations only for event type $A$. 
In the table, $F_A$ and $F_B$ represent the  frequency of event types $A$ and $B$ in the type frequency $F$ in the predecessor pane $\omega^e$.

\gSPICE{} aggregates observations that have the same values for event types $T_e$, type frequency $F$ in the predecessor pane, and event attributes $\mathbb{E}_e$ into  a set of aggregated observations (denoted by $\mathbb{S}_g$).
\sloppy An observation $ob_g \in \mathbb{S}_g$ is of the following form: $ob_g \langle T_e, F, \mathbb{E}_e, M, O \rangle \in \mathbb{S}_g$. $M$ corresponds to the occurrences of complex events in the set $\mathbb{M}_e$ in observations $ob_e \langle T_e, \mathbb{F}_{e}, \mathbb{E}_e, *\rangle \in \mathbb{S}_e$, where  $M$  represents the sum of the occurrences of the complex events in $\mathbb{M}_e$ multiplied by their weights, as complex events have weights reflecting their importance.
$O$ represents the number of occurrences of these observations $ob_e \langle T_e, \mathbb{F}_{e}, \mathbb{E}_e, *\rangle \in \mathbb{S}_e$. The sign $*$ is used as a wildcard for the set of complex events $\mathbb{M}_e$ in the observations $ob_e \in \mathbb{S}_e$. The following equation formally formulates  how \gSPICE{} builds the aggregated observations.
\begin{align} 
\begin{split}
 \textstyle ob_g &\langle T_e, F, \mathbb{E}_e, M, O \rangle  \in \mathbb{S}_g: \\
 &  \textstyle M =  \textstyle \sum_{ob_e \langle T_e, F, \mathbb{E}_e, \mathbb{M}_e\rangle \in \mathbb{S}_e} {\sum_{cplx \in \mathbb{M}_e}{ w_{q_i}: cplx \subset q_i}} \\[0.2cm]
 &  \textstyle O = \textstyle \sum _{ob_e \langle T_e, F, \mathbb{E}_e, \mathbb{M}_e\rangle \in \mathbb{S}_e} {1}
\end{split}
\label{eq:obs}
\end{align} 
In Figure \ref{fig:observations-example}, Table \ref{tab:aggregated-obsevations} shows the  aggregated observations as a result of aggregating observations in Table \ref{tab:obsevations}. For example, in Table \ref{tab:obsevations}, an event $e$ of type $T_e= A$ with a type frequency $F= (F_A, F_B)= (1, 2)$ and event attribute $E_e= 5$ occurs three times and contributes two times to a single complex event (i.e., $cplx_1$ and $cplx_2$). Hence, $O=3$ and $M= 2$,  where $cplx_1 \subset q$ and $cplx_2 \subset q$, assuming that the pattern's weight $w_q=1$. That results in the following aggregated observation in Table \ref{tab:aggregated-obsevations}: $ob_g \langle A, (1, 2), \{5\}, 2, 3\rangle$. 

Please note that, as we mentioned above, if a query  contains the \text{negation} event operator, a PM is abandoned when the negated event matches the  pattern. Hence, in this case, no complex events are detected. Therefore, to capture the importance of the negated events, we assume that the CEP operator forwards the abandoned PMs to \gSPICE{} to learn about the utility of these negated events.

\definecolor{LightCyan}{rgb}{0.88,1,1}
\newcolumntype{g}{>{\columncolor{LightCyan}}c}
\renewcommand{\arraystretch}{1}
\setlength{\fboxrule}{1pt}
\begin{figure}
	\resizebox{0.90\linewidth}{!}{	
		\centering
		{\fontsize{10.0}{11.0}\selectfont
		%	\fbox{%
		\begin{minipage}[t]{\linewidth}	
			\centering	
			\begin{minipage}[t]{0.99\linewidth}
				\centering
				\begin{tabular}[]{| c | c | c |c| c|c| c|}
					%				\specialrule{.10em}{0em}{0em} 		
					\hline%
					\rowcolor{LightCyan}
					$T_e$ & $F_A$ &$F_B$ & $E_e$ & $\mathbb{M}_e$ \\ \hline%   %\cline{2-2} 
					$A$ & $1$ &$2$ & $5$ & $\{cplx_1\}$ \\ \hline%   %\cline{2-2} 
					$A$ & $1$ &$2$ & $5$ & $\{\}$ \\ \hline%   %\cline{2-2} 
					$A$ & $1$ &$2$ & $5$ & $\{cplx_2\}$ \\ \hline%   %\cline{2-2} 
					$A$ & $2$ &$1$ & $7$ & $\{\}$ \\ \hline%   %\cline{2-2} 
					$A$ & $2$ &$1$ & $7$ & $\{cplx_3\}$ \\ \hline%   %\cline{2-2} 
					$A$ & $2$ &$1$ & $8$ & $\{\}$ \\ \hline%   %\cline{2-2} 
					$A$ & $2$ &$1$ & $8$ & $\{\}$ \\ \hline%   %\cline{2-2} \\
					$A$ & $2$ &$1$ & $8$ & $\{cplx_4\}$ \\ \hline%   %\cline{2-2} \\
					
					%				\specialrule{.10em}{0em}{0em} 
				\end{tabular}
				\captionof{table}{Observations $\mathbb{S}_e$.}
				\label{tab:obsevations}
			\end{minipage} 
			\hfill \vspace{0.2cm} \hfil%
			\begin{minipage}[t]{0.99\linewidth}
				\centering
				\begin{tabular}[]{| c | c | c |c| c|c| c| c| c|}
					%				\specialrule{.10em}{0em}{0em} 		
					\hline%
					\rowcolor{LightCyan}
					$T_e$ & $F_A$ &$F_B$ & $E_e$ & $M$ & $O$ & $U_e$\\ \hline%   %\cline{2-2} 
					$A$ & $1$ &$2$ & $5$ & $2$ & $3$ & $0.67$\\ \hline%   %\cline{2-2} 
					$A$ & $2$ &$1$ & $7$ & $1$ & $2$ & $0.50$\\ \hline%   %\cline{2-2} 
					$A$ & $2$ &$1$ & $8$ & $3$ & $1$ & $0.33$\\ \hline%   %\cline{2-2} 							
					%				\specialrule{.10em}{0em}{0em} 
				\end{tabular}
				\captionof{table}{Aggregated Observations $\mathbb{S}_g$ and the predicted utilities.}
				\label{tab:aggregated-obsevations}
			\end{minipage}
		\end{minipage}	
		%	}
	}
	}	
%	\vspace{-0.6cm}
	\caption{Statistic gathering and utility calculation.}
	\label{fig:observations-example}
%	\vspace{-0.6cm}
\end{figure}

\subsubsection{Utility Prediction}
After gathering statistics from $\eta$ observations, \gSPICE{} uses these observations to predict the utility function $f$, hence the event utility $U_e$. Equation \ref{eq:u_ob} shows the way \gSPICE{} computes the event utility $U_e$ from the aggregated observations:
\begin{equation} 
U_e =f(T_e, F, \mathbb{E}_e)=  \textstyle \frac{ M}{O} : \forall ~ ob_g \langle T_e, F, \mathbb{E}_e, M, O \rangle  \in \mathbb{S}_g
\label{eq:u_ob}
\end{equation}
To compute the utility $U_e$ of an event $e$ of type $T_e$,  with type frequency $F$, and event attributes $\mathbb{E}_e$, for an aggregated observation $ob_g \langle T_e, F, \mathbb{E}_e, M, O \rangle \in \mathbb{S}_g $, \gSPICE{} divides $M$ in the aggregated observation $ob_g$ by the number of occurrences  $O$ in this aggregated observation $ob_g$. 
Table \ref{tab:aggregated-obsevations} also shows the computed utilities from the aggregated observations.  For example, in the table, in the aggregated observation $ob_g \langle A, (1, 2), \{5\}, 2, 3\rangle$, $M= 2$ and $O= 3$. Therefore, the utility $U_e$ of an event $e$ of type $T_e= A$ with a type frequency $F= (F_A,F_B)= (1,2)$ and event attribute $E_e= 5$ is calculated as follows:  $U_e= \frac{2}{3}= 0.67$.

The distribution of events in the input event stream may change over time, where the predicted event utilities might become inaccurate. To keep predicted event utilities accurate, they may be either periodically recomputed or only when the distribution of events in the input event stream changes by a certain threshold.  

To use the predicted event utilities $U_e$ during load shedding,  \gSPICE{} handles the predicted utilities in the following two ways. 1)  \gSPICE{} stores the utilities in hash tables. We refer to this approach as \gSPICESH. 2) \gSPICE{} trains a well-known machine learning model (e.g., a decision tree  or a random forest) with the predicted utilities to estimate the utility function. We refer to this approach as \gSPICESM.

\subsubsection{\gSPICESH}
In \gSPICESH, we store the  event utilities in a utility table (denoted by $UT$). The data structure used to store the utility table $UT$ consists of hash tables. For each event type $T_e$, there is a hash table that stores the event utilities for all observed combinations of the type frequency $F$ in the predecessor pane $\omega^e$ and event attributes $\mathbb{E}_e$. Hence, the utility of an event $e$ of type $T_e$ is stored in the utility table as follows: $U_e = f(T_e, F, \mathbb{E}_e)= UT[T_e][K]$ where the hash key $K$ is computed from the type frequency $F$ and event attributes $\mathbb{E}_e$. 
The values of event attributes $\mathbb{E}_e$ might occupy a wide range. Similarly,  an event type $T_e$ might have a value between zero and  $L_\omega$ in the type frequency $F$. That might result in a huge number of combinations of different event attributes $\mathbb{E}_e$  and type frequency $F$, especially if the length $L_\omega$ of the predecessor pane is large. That might considerably increase the required memory to store the utilities. To reduce the needed memory to store the utility table $UT$, we group the successive values of event attributes and the successive frequencies in the type frequency using bins of fixed sizes \cite{Kotsiantis2007DataPF}. To simplify the presentation, we assume that a bin of size one is used for event attributes and type frequency if not otherwise stated.

To get the hash key $K$, we need to implement a hash function that combines the type frequency and event attributes. 
One way to implement the hash function is by using a function that iterates over event types in the type frequency and over the event attributes to compute the hash key $K$. The computational overhead of this approach depends on the number of event types in the type frequency and on the number of event attributes.  The number of event types might be high, hence the computational overhead to get the key $K$ might be high. To reduce the overhead, we use the Zobrist hashing \cite{Zobrist1990ANH} as a hashing function to compute the key $K$.

Zobrist hashing depends on bitwise XOR operations (denoted by $\oplus$)  to compute the hash key \cite{Zobrist1990ANH}. 
To use the Zobrist hashing, we do the following. 
1) We generate big unique random numbers for each possible frequency of an event type in the type frequency $F$. Hence, for each event type $T_e$, we generate the following set $\mathbb{R}_{T_e}$ of random numbers: $\mathbb{R}_{T_e}= \{ R^{l}:  0 \le l \le L_\omega \} $ where $R^{l}$ represents a big unique random number. Please note that an event type $T_e$ might at most occur $L_\omega$ times in the type frequency.  
2) We also generate big unique random numbers for each possible range of each event attribute.  Hence, for an event attribute $E_e \in \mathbb{E}_e$, we generate the following set $\mathbb{R}_{E_e}$ of random numbers: $\mathbb{R}_{E_e}= \{ R^{l}:  0 \le l \le max(range) \} $ where $R^{l}$ represents  a big unique random number and $max(range)$ represents the maximum range an event attribute might have. 
3) We use one hash key $K_1$ as a hash key for event types in the type frequency. The hash key $K_1$ is computed as follows: 
$K_1= \bigoplus R_{T_i}^{F_{T_i}} : \forall~ T_i \in \mathbb{T}$. The key $K_1$ is computed by performing XOR operations between the random numbers corresponding to the frequency of each event type in the type frequency.
4) We use a second hash key $K_2$ as a hash key for event attributes. The hash key $K_2$ is computed as follows: 
$K_2= \bigoplus R_{E_e}^l : \forall~ E_e \in \mathbb{E}_e, l = range_E$,  where $range_E$ represents the corresponding range for the value of the event attribute $E_e$. Similar to computing $K_1$, the key $K_2$ is computed by performing XOR operations between the random numbers corresponding to the range values of each event attribute. 
5) The hash key $K$ is computed by performing an XOR operation between the hash keys $K_1$ and $K_2$, i.e., $K= K_1 \oplus K_2$.	 

To reduce the overhead of computing the hash key $K$, we \textit{continuously update} the hash key $K_1$ which is computed from event types in the type frequency. We first compute the hash key $K_1$ for all event types in the type frequency. Then, when the predecessor pane changes, for all changed frequencies of event types in the type frequency,  we remove the old values and add the new values to the hash key. Since XOR is a self-inverse operation (e.g., $X \oplus X= 0$), to remove an old value $F_{T_i}^\prime$  for the event type $T_i$ in the type frequency, we need to perform only a single XOR operation. Additionally, we need a single XOR operation to add the new value $F_{T_i}$ for the event type $T_i$ in the type frequency. Hence, for each changed frequency of an event type, we need two XOR operations, i.e., $K_1= K_1 \oplus R_{T_i}^{F_{T_i}^\prime} \oplus R_{T_i}^{F_{T_i}}$. Therefore, if the predecessor pane changes by only one event (i.e., the predecessor pane shifts by one event), there is a need for a maximum of four XOR operations.
%to update the hash key $K_1$.  
%That is because an event type might be removed from the predecessor pane (i.e., needs two XOR operations) and another event type might be added to the predecessor pane (i.e, needs two XOR operations). 
As a result, to compute a new hash key $K$, there is a need for at most (4 + $|\mathbb{E}_e |$) XOR operations, resulting in a considerable reduction in the overhead of computing the hash key $K$, especially, in the case of a high number of event types.

\subsubsection{\gSPICESM}
Another way to handle a huge number of predicted utilities is to use well-known machine learning models where we might use a machine learning model to estimate the utility function. In this case, the aggregated observations $\mathbb{S}_g$ are used  as input training data to the machine learning model, and the corresponding computed utility values (cf. Equation \ref{eq:u_ob}) are used as labels. After training the model, the produced trained model represents the estimated utility function. Hence, to get the utility of an event $e$, \gSPICESM{} provides the trained model with the event type $T_e$, type frequency $F$ in the predecessor pane, and event attributes $\mathbb{E}_e$. The model returns the predicted utility value $U_e$.

Several machine learning models can be used to estimate the utility function, e.g., neural networks, decision trees, random forest, etc. 
However, machine learning models usually impose considerable computational overhead that might not be tolerable for performing load shedding in CEP. Moreover, these models have many parameters that need to be tuned, which increases the burden on a domain expert.
%For example, using neural networks  with  many layers might introduce a high number of parameters. Besides that, it imposes a high prediction overhead. 
In \gSPICESM, we use two machine learning models, namely, decision trees \cite{10.5555/152181, Anyanwu2009ComparativeAO} and random forests \cite{598994}, to estimate the utility function. Please note that controlling the depth of trees in these models and other parameters, such as when to split nodes, can control  the needed memory size by these two models. In Section \ref{sec:results}, we show the performance of these models in estimating the utility function and their imposed computational overhead.

\subsection{Utility Threshold}
As we mentioned above, in overload cases, \gSPICE{} must drop $\rho\%$ of events during every drop interval (e.g., window) to maintain the given latency bound. To do that, \gSPICE{} uses the predicted event utilities to find a utility threshold $u_{th}$ that can be used to drop the required percentage of events $\rho$. That is done in a similar way to predicting utility threshold in \cite{espice}. First, \gSPICE{} uses the gathered statistics and the predicted event utilities to compute the percentage of occurrences (denoted by $O_u$) of an event utility $u$ in the already processed event stream $\mathbb{S}_p$ in the gathered statistics, i.e., $O_u= \frac{ |\{e: e\in \mathbb{S}_p, u=U_e\}|} {|\mathbb{S}_p|}$. Then, \gSPICE{} accumulates the percentage of occurrences of event utilities $O_u$ in ascending order to get the cumulative utility occurrences (denoted by $O_u^c$) as follows: $O_u^c = \sum_{u^\prime \le u} {O_{u^\prime}}$. The cumulative utility occurrences $O_u^c$ represents the percentage of events in the stream of already processed events $\mathbb{S}_p$ that have a utility value $u^\prime$ which is less than or equal  to the utility $u$, i.e., $u \le u^\prime$.  Hence, to drop $\rho\%$ of events from every drop interval, we may find the lowest cumulative utility occurrences $O_u^c$ that is higher than or equal to $\rho$, and choose the utility $u$ as the utility threshold $u_{th}$, i.e., $u_{th}= u$: $O_u^c \ge \rho ~\forall~ O_{u^\prime}^c \ge \rho \land  O_u^c \le O_{u^\prime}^c$. 

\subsection{Load Shedding}
%After we have explained how to predict the event utility, handle the predicted utilities, and then predict utility threshold, now,
In this section, we explain the way \gSPICE{} uses  predicted event utilities and  utility thresholds to drop events from the input event stream $S_{in}$ during overload to maintain a given latency bound. 
Algorithm \ref{alg:loadShedding} formally defines how load shedding is performed.  

For each event $e$ in the input event stream $S_{in}$, before event $e$ is processed by the operator, \gSPICE{} checks whether there is overload and a need to drop events. If there is no overload, the event is processed by the operator (cf. Algorithm \ref{alg:loadShedding}, lines 2-3). Otherwise, the operator is overloaded and there is a need to drop events. In this case, \gSPICE{} must drop $\rho\%$ of events from the input event stream $S_{in}$ in every drop interval to maintain the given latency bound. Therefore, \gSPICE{} first finds a utility threshold $u_{th}$ that results in dropping $\rho\%$ of events using the cumulative utility occurrences as we explained above. In case the event utilities are stored using the Zobrist hashing, i.e., \gSPICESH (cf. Algorithm \ref{alg:loadShedding}, lines 4-10), \gSPICE{} computes the hash key $K$ by XORing the hash key $K_1$ for the type frequency $F$ and the hash key $K_2$ for the event attributes $\mathbb{E}_e$  (cf. Algorithm \ref{alg:loadShedding}, lines 5-6). Then, \gSPICE{} gets the event utility $U_e$ from the utility table $UT$ and compares the utility $U_e$ with the utility threshold $u_{th}$. If the event utility $U_e$ is less than or equal to the utility threshold, the event $e$ is dropped from the input event stream $S_{in}$ (cf. Algorithm \ref{alg:loadShedding}, lines 7-8). Otherwise, the event $e$ is processed normally by the operator. In case \gSPICESM{} is used to estimate the event utility (cf. Algorithm \ref{alg:loadShedding}, lines 11-15), \gSPICE{} provides the model with the event type, type frequency, and event attributes. The model returns the predicted event utility $U_e$. Here again, if the event utility $U_e$ is less than or equal to the utility threshold $u_{th}$, the event $e$ is dropped from the input event stream $S_{in}$ (cf. Algorithm \ref{alg:loadShedding}, lines 12-13). Otherwise, event $e$ is processed by the operator.

\begin{algorithm}
	\setbox0\vbox{\small
		{\fontsize{9.0}{10.0}\selectfont
			\begin{algorithmic}[1]
				\algsetblockdefx[function]{func}{endfunc}{}{0.2cm}[3]{#1 \textbf{#2} (#3) \textbf{begin}}{\textbf{end function}}				
				
				\func {}{drop}{$e$}
				% {\fontsize{7.0}{8.0}\selectfont \Comment $T_e$: index of the event type, $P_e$: event position in a widow, and $S_{\gamma}$: index of PM state.}			
				\If{$!\mathit{isOverloaded}$} {\fontsize{8.0}{9.0}\selectfont \Comment no overload hence no need to drop events}	
					\State $\mathbf{return} \quad False$
				
				\ElsIf {$\mathit{isZobrist}$} {\fontsize{8.0}{9.0}\selectfont \Comment using \gSPICESH.}
					\State $K_2= \bigoplus \mathbb{E}_e$ {\fontsize{7.0}{8.0}\selectfont \Comment computing $K_2$ by XORing event attributes $\mathbb{E}_e$.}
					\State $K= K_1 \oplus K_2$ {\fontsize{8.0}{9.0}\selectfont \Comment $K_1$ is continuously updated.}
					\If {$UT[T_e][K] \le \mathit{u_{th}}$}
						\State $\mathbf{return} \quad True$
					\Else
						\State $\mathbf{return} \quad False$	
					\EndIf		
				\Else {\fontsize{7.0}{8.0}\selectfont \Comment using \gSPICESM.}
					\If {$\mathit{model.getUtility}(T_e, F, \mathbb{E}_e) \le \mathit{u_{th}}$}
						\State $\mathbf{return} \quad True$
					\Else
						\State $\mathbf{return} \quad False$	
					\EndIf		
					
				\EndIf		
				
				\endfunc
				
			\end{algorithmic}
		}
	}
	\centerline{\fbox{\box0}}
%	\vspace{-0.4cm}
	\caption{Load shedder.}
	\label{alg:loadShedding}
\end{algorithm}
%\vspace{-0.6cm}

\section{Performance Evaluations}
\label{sec:results}
In this section, we evaluate the performance of \gSPICE{} by using several datasets and a set of representative queries.

\subsection{Experimental Setup}
\label{sec:experimental-setup}
Here, we describe the evaluation platform, the baseline implementation, datasets, and queries used in the evaluations.

\textbf{\textit{Evaluation Platform.}}
We run our evaluation on a machine that is equipped with 8 CPU cores (Intel 1.6 GHz) and a main memory of 24 GB. The  OS used is CentOS 6.4. We run the operator in a single thread that is used as a resource limitation. Please note, that the performance of \gSPICE{} is independent of the parallelism degree of the operator.
We implemented \gSPICE{} by extending a prototype CEP framework which is implemented using Java.

\textbf{\textit{Baseline.}}
We compare the performance of \gSPICE{} with two state-of-the-art black-box load shedding strategies: 1) eSPICE: a shedding approach that drops events from windows \cite{espice}, where an event is assigned a utility depending on its type and position in a window. 2) BL: a load shedding strategy similar to the one proposed in \cite{He2014OnLS}. Our implementation also captures the notion of weighted sampling techniques in stream processing \cite{Tatbul:2003:LSD:1315451.1315479}. BL drops events from windows, where an event type receives a higher utility proportional
to its repetition in patterns and windows. Then, depending on event type utilities, it uses uniform sampling to decide which event instances to drop from the same event type. 

\textbf{\textit{Datasets.}}
The distribution of event types might considerably impact the performance of \gSPICE. Therefore, to control the event distribution, we generate eight synthetic datasets as shown in Table \ref{tab:synthetic-datasets}, where events are generated using an exponential distribution. Datasets  $DS_1$, $DS_2$, $DS_3$, and $DS_4$ contain events of three types: $A$, $B$, and $C$. While datasets $DS_5$, $DS_6$, $DS_7$, and $DS_8$ contain events of six types: $A$, $B$, $C$, $D$, $E$, and $F$. $\mu_X$ represents the average time (in seconds) between event instances of the event type $X$. $\mu_X$ controls the percentage of each event type in these datasets. Table \ref{tab:synthetic-datasets} shows the average expected percentage (approximate values) of each event type in the datasets. Moreover, all events in all datasets have an  attribute $V_1$ with uniformly distributed values between 1 and 10.
We also use two real-world datasets. 1) A stock quote stream (denoted by NYSE dataset) from the New York Stock Exchange (NYSE), which contains real intra-day quotes of different stocks from NYSE collected over two months from Google Finance \cite{google_finance}. This dataset contains stock events that have a change in their quote by at least 0.4\%. 
2) A position data stream from a real-time locating system (denoted by RTLS dataset) in a soccer game \cite{debs2013}. Players, balls, and referees  are equipped with sensors that periodically generate events containing their position, velocity, etc.

% define query names
\newcommand{\syntheticQOne}{$Q_1$}
\newcommand{\syntheticQTwo}{$Q_2$}
\newcommand{\syntheticQFour}{$Q_3$}
\newcommand{\stockQOne}{$Q_4$}
\newcommand{\stockQThree}{$Q_5$}
\newcommand{\soccerQOne}{$Q_6$}
%end query name definition

%resize figrues
\newcommand{\figureWidthFourInRow}{0.22}
\newcommand{\figureWidthTwoInRow}{0.44}
\newcommand{\figureWidthOneInRow}{0.44}

\textbf{\textit{Queries.}}
We employ six queries that cover an important set of operators in CEP as shown in Table \ref{tab:queries}: sequence, disjunction,  sequence with Kleene closure, sequence with negation, and sequence with any operators  \cite{ch1994snoop, cu2010tesla, Wu:2006:HCE:1142473.1142520}. Queries \syntheticQOne, \syntheticQTwo, and \syntheticQFour{} are executed over the synthetic datasets. While queries \stockQOne{} and \stockQThree{} are executed over the NYSE dataset. \soccerQOne{} is executed over the RTLS dataset.
We use the first selection policy for all events in all queries. Moreover, we use the consumed consumption policy for the first event in all queries and the zero consumption policy for the rest of the events in all queries \cite{ch1994snoop, Zimmer:1999:SCE:846218.847253}. 
In  the table, $C_i$ represents the stock quote of company $i$, and $D_i$ represents the event of player $i$.

\definecolor{LightCyan}{rgb}{0.88,1,1}
\newcolumntype{g}{>{\columncolor{LightCyan}}c}
\newcolumntype{?}[1]{!{\vrule width #1}}
\definecolor{Gray}{gray}{0.95}
\renewcommand{\arraystretch}{1}
\setlength{\fboxrule}{1pt}
\begin{table}[h]
	\resizebox{0.99\linewidth}{!}{%
		\begin{tabular}{| g | c| c | c |c |c |c ?{0.5mm}c | c | c |c |c |c |}
			%		\hline 
			%		\cline{1-7}
			\hhline{~*{12}{-}}
			\rowcolor{LightCyan}
			\multicolumn{1}{g|}{\cellcolor{white}}  & \multicolumn{1}{c|}{ $\mu_A$} & $\mu_B$ & $\mu_C$ & $\mu_D$ & $\mu_E$ & $\mu_F$ & $A\%$ & $B\%$ & $C\%$ & $D\%$ & $E\%$  & $F\%$ \\ \hline			   
			$DS_1$ & 2.5 & 15 & 40 & -   & -  & -  & 81.3 & 13.6 & 5.1  & -  & -  & -  \\ \hline
			$DS_2$ & 2.8 & 15 & 15 & -   & -  & -  & 72.8 & 13.6 & 13.6 & -  & -  & - \\ \hline
			$DS_3$ & 4   & 6  & 12 & -   & -  & -  & 50   & 33.3 & 16.7 & -  & -  & - \\ \hline
			$DS_4$ & 6   & 6  & 6  & -   & -  & -  & 33.3 & 33.3 & 33.3  & -  & -  &- \\ \hline
			$DS_5$ & 2.5 & 15 & 40 & 2.5 & 15 & 40 & 40.7 & 6.8 & 2.5  & 40.7  & 6.8  & 2.5  \\ \hline
			$DS_6$ & 2.8 & 15 & 15 & 2.8 & 15 & 15 & 36.4 & 6.8 & 6.8 & 36.4  & 6.8  & 6.8 \\ \hline
			$DS_7$ & 4   & 6  & 12 & 4   & 6  & 12 & 25   & 16.7 & 8.3 & 25  & 16.7  & 8.3 \\ \hline
			$DS_8$ & 6   & 6  & 6  & 6   & 6  & 6  & 16.7 & 16.7 & 16.7  & 16.7  & 16.7  & 16.7 \\ \hline
			
		\end{tabular}
	}
%		\vspace{-0.6cm}
	\caption{Synthetic Datasets.}
	\label{tab:synthetic-datasets}	
	\vspace{-0.5cm}
\end{table}

\renewcommand{\arraystretch}{1}

\begin{table}[t]
	\centering
	\resizebox{.99\linewidth}{!}{%	
		\begin{tabular}{| l | l |}
			\noalign{\hrule height 1pt}
			%	\rowcolor{LightCyan}
			\multicolumn{2}{!{\vrule width 1pt}c!{\vrule width 1pt}}{\textbf{Queries on synthetic data}} \\
			\noalign{\hrule height 1pt}		
			\syntheticQOne{}&  \makecell[cl]{\textbf{pattern} $\mathbf{seq} (A; B; C)$ \\
				\qquad $\mathbf{where}~  A.V_1 < B.V_1 ~ $  
				$\textbf{and}~ A.V_1 + B.V_1 < C.V_1$\\
				\qquad\textbf{within} \textit{ws} \text{seconds} 
			}
			\\ \hline
			
			\syntheticQTwo{}&  \makecell[cl]{\textbf{pattern} $\mathbf{seq} (A;B;C) \vee \mathbf{seq} (D;E;F)$\\
				\qquad $\mathbf{where}~  (A.V_1 < B.V_1 ~ $  
				$\textbf{and}~ A.V_1 + B.V1 < C.V_1)$\\
				\qquad $\textbf{or}~  (D.V_1 < E.V_1 ~ $  
				$\textbf{and}~ D.V_1 + E.V1 < F.V_1)$\\				
				\qquad\textbf{within} \textit{ws} \text{seconds} 	 
			}\\ \hline
			
%			\syntheticQThree{}&  \makecell[cl]{\textbf{pattern}  $\mathbf{seq}(A; B; C)$   \\                
%			\qquad $\mathbf{where}~ (A.V_1 > A.V_2 ~$
%			$\textbf{and}~ A.V_1 + B.V_1 >  A.V_2 + B.V_2$ \\
%			\qquad  $\textbf{and}~ A.V_1 + B.V_1 + C.V_1 >  A.V_2 + B.V_2 + C.V_2)$ \\
%			\qquad $\textbf{or}~ (A.V_1 \le A.V_2 ~$
%			$\textbf{and}~ A.V_1 + B.V_1 \le  A.V_2 + B.V_2$ \\
%			\qquad  $\textbf{and}~ A.V_1 + B.V_1 + C.V_1 \le A.V_2 + B.V_2 + C.V_2)$ \\
%			\qquad\textbf{within} \textit{ws} \text{seconds} 
%			}\\   \hline
%		
			\syntheticQFour{}&  \makecell[cl]{\textbf{pattern} $\mathbf{seq} (A; B+; C)$ \\
			\qquad $\mathbf{where}~  A.V_1 + \sum_{i < j}{B_i.V_1} < B_j.V_1 ~ $  
			$\textbf{and}~ A.V_1 + \sum {B.V_1} < C.V_1$\\
			\qquad\textbf{within} \textit{ws} \text{seconds} 
			}	\\ %\hline
	
%		\syntheticQSix &  \makecell[cl]{\textbf{pattern} $\mathbf{any} (3, A; B; C)$ \\
%		\qquad $\mathbf{where}~  A.V_1 < B.V_1 ~ $  
%		$\textbf{and}~ A.V_1 + B.V_1 < C.V_1$\\
%		\qquad\textbf{within} \textit{ws} \text{seconds} 
%		}	\\ 
			
			%%%%%%%% stock %%%%%%%%
		\noalign{\hrule height 1pt}
		\multicolumn{2}{!{\vrule width 1pt}c!{\vrule width 1pt}}{\textbf{Stock queries}} \\
		\noalign{\hrule height 1pt}
		\stockQOne &  \makecell[cl]{\textbf{pattern} $\mathbf{seq} (C_1; C_2;..;C_{10})$ \\
			\qquad $\mathbf{where}~ all~ C_i ~rise~ by~ x\%$ 
			$\textbf{or} ~ all~ C_i ~fall ~ by~ x\%,~ i= 1.. 10$\\
			\qquad\textbf{within} \textit{ws} \text{minutes} 
		}
		\\ \hline
		
%		\stockQTwo &  \makecell[cl]{\textbf{pattern} $\mathbf{seq} (C_1;  C_1; C_2; C_3; C_2; C_4; C_2;$
%			$C_5; C_6; C_7; C_2; C_8; C_9; C_{10})$\\
%			\qquad $\mathbf{where}~ all~ C_i ~rise~ by~ x\%$ 
%			$\textbf{or} ~ all~ C_i ~fall~ by~ x\%,~ i= 1.. 10$\\
%			\qquad\textbf{within} \textit{ws} \text{minutes} 	 
%		}\\ \hline
%	
		\stockQThree &  \makecell[cl]{\textbf{pattern}  $\mathbf{seq}(C_1;  C_2; C_3; C_4; \mathbf{!C_5}; C_6; C_7; C_8; C_9; C_{10})$   \\                
			\qquad $\mathbf{where}~ all~ C_i ~rise~ by~ x\% ~\mathbf{and}~ C_5 ~does~not~ rise~ by~ y\%$ \\
			\qquad$\textbf{or} ~ all~ C_i ~fall~ by~ x\% ~\mathbf{and}~ C_5 ~does~not~ fall~ by~ y\%$\\
			\qquad \quad $,~ i= 1.. 10 ~ and~ i\ne 5$\\
			\qquad\textbf{within} \textit{ws} \text{minutes} 
		}\\ 
	
			%%%%%%%% soccer %%%%%%%%
		\noalign{\hrule height 1pt}
		\multicolumn{2}{!{\vrule width 1pt}c!{\vrule width 1pt}}{\textbf{Soccer queries}} \\ 
		\noalign{\hrule height 1pt}
		\soccerQOne &   \makecell[cl]{\textbf{pattern} $\mathbf{seq} (S; \mathbf{any}(3, D_1, D_2, ..,D_m))$ \\
			\qquad $\mathbf{where}~ S ~possesses ~ball~ \mathbf{and} ~ distance(S, D_i) \leq x~ meters $ \\
			\qquad \qquad $,~ i= 1.. m$  and $n$ is the number of players in a team\\
			\qquad\textbf{within} $ws$ \text{seconds}
		}		
		\\
			
		\noalign{\hrule height 1pt}	
			
		\end{tabular}
	}
%	\vspace{-0.4cm}
	\captionof{table}{Queries.}
	\label{tab:queries}
	\vspace{-0.5cm}
\end{table}

\subsection{Experimental Results}
We evaluate the impact of \gSPICE{} on QoR, particularly on the number of false positives and false negatives, and compare its results with that of BL and eSPICE.  Moreover, we show the overhead of \gSPICE{} and its ability to maintain a given latency bound (LB). 
%In our evaluations, we observed that dropping events on the stream granularity in \gSPICE{} always results in  better performance, w.r.t. QoR, compared to dropping events on the window granularity. Therefore, in this section, we only show results for \gSPICE{} when dropping events on the stream granularity. 
We first focus on the performance of \gSPICESH, i.e., when Zobrist hashing is used.  Later, we also show the evaluation results of \gSPICE{} when using a decision tree or a random forest. 

If not noted otherwise, we employ the following settings. For all queries, we use a time-based sliding window and a time-based predicate. For queries based on synthetic data (i.e.,  \syntheticQOne, \syntheticQTwo, and \syntheticQFour), we use a window of size 250 seconds. For \syntheticQOne{} and \syntheticQFour, a  new window is opened every 10 seconds (i.e., the slide size is 10 seconds). While for \syntheticQTwo, a  new window is opened every 20 seconds. For stock queries (i.e., \stockQOne{} and \stockQThree), we use a window of size 15 minutes and a slide of size 1 minute. Query \soccerQOne{} (i.e., the soccer query) uses a window of size 30 seconds and a slide of size 1 second.
We stream events to the operator from the datasets stored in files. We first stream events at input event rates that are less or equal to the operator throughput $\mu$ (maximum service rate) until the model is built. After that, we increase the input event rate to enforce load shedding, as we will mention in the following experiments. The used latency bound $LB = 1$ second. We configure all load shedding strategies to have a safety bound, where they start dropping events when the event queuing latency is greater or equal to 80\% of LB, i.e., the safety bound equals 200 milliseconds. We execute several runs for each experiment and show the mean value and standard deviation.

Several factors influence the performance of \gSPICE{}, e.g., event rate, event distribution, and predecessor pane length $L_\omega$. Therefore, we analyze the performance of \gSPICE{} with these different factors.

\subsubsection{Results on Synthetic Data}
\label{sec:impact-of-event-rate}
To evaluate the performance of \gSPICE, we run experiments with queries \syntheticQOne, \syntheticQTwo, and \syntheticQFour{} with event rates 120\%, 140\%, 160\%, 180\%, and 200\% of the operator throughput $\mu$ (i.e., the input event rate is higher than the operator throughput $\mu$ by 20\%, 40\%, 60\%, 80\%, and 100\%). We use the dataset $DS_1$ for  \syntheticQOne{} and  \syntheticQFour{}  and the dataset $DS_5$  for  \syntheticQTwo. For all queries, we use a predecessor pane of length 10 events, i.e., $L_\omega= 10$.

\begin{figure*}[t]
	\centering
	\begin{subfigure}[t]{\figureWidthFourInRow\linewidth}
		\includegraphics[width=\linewidth]{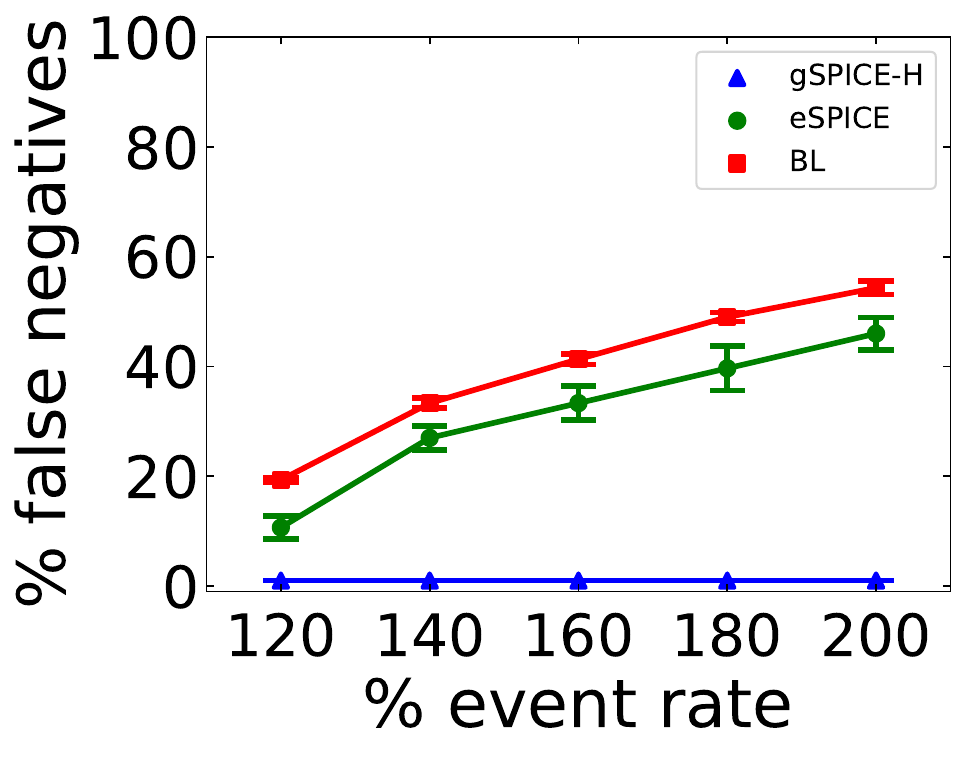} 
%		\vspace{-0.6cm}
		\caption[t]{\syntheticQOne: false negatives}
		\label{fig:synthetic-q1-strict-fn-r}
	\end{subfigure}
	\hfill%
	\begin{subfigure}[t]{\figureWidthFourInRow\linewidth}
		\includegraphics[width=\linewidth]{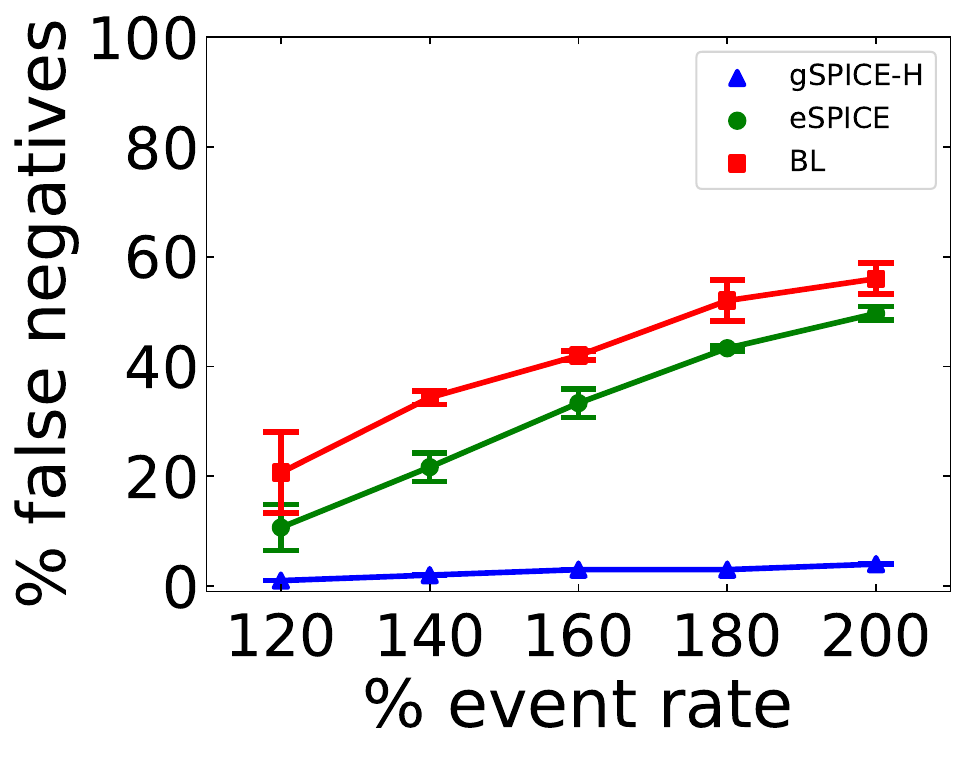}
%		\vspace{-0.6cm}
		\caption[]{\syntheticQTwo: false negatives}
		\label{fig:synthetic-q2-strict-fn-r}
	\end{subfigure}
	\hfill%
	\begin{subfigure}[t]{\figureWidthFourInRow\linewidth}
		\includegraphics[width=\linewidth]{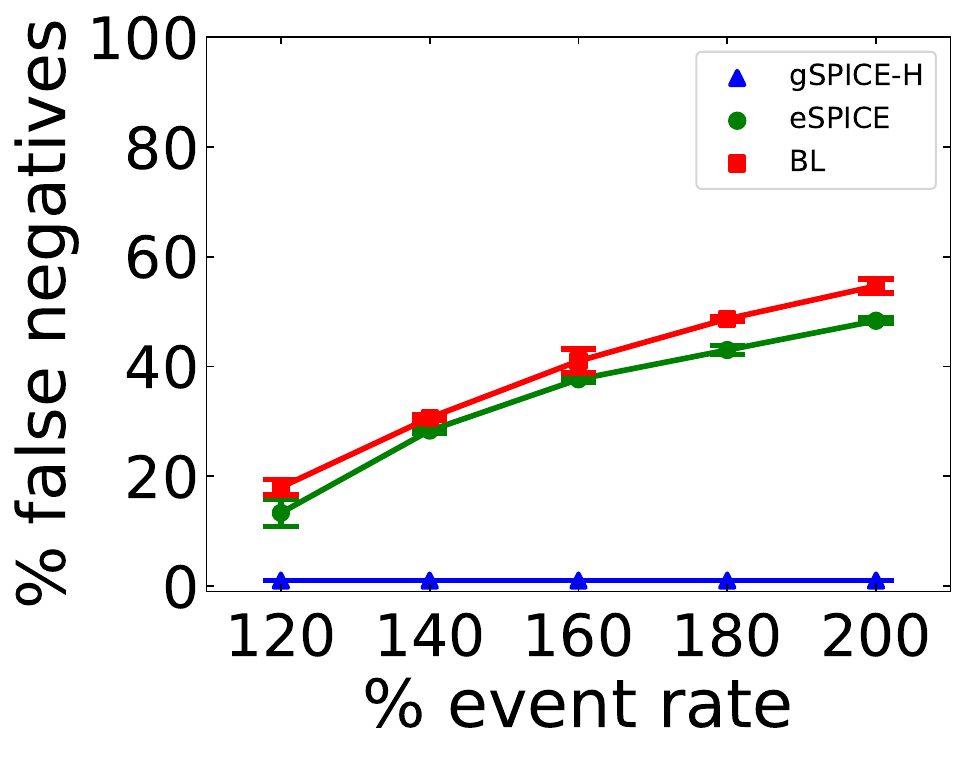} 
%		\vspace{-0.6cm}
		\caption[t]{\syntheticQFour: false negatives}
		\label{fig:synthetic-q4-strict-fn-r}
	\end{subfigure}
	\hfill
	\begin{subfigure}[t]{\figureWidthFourInRow\linewidth}
		\includegraphics[width=\linewidth]{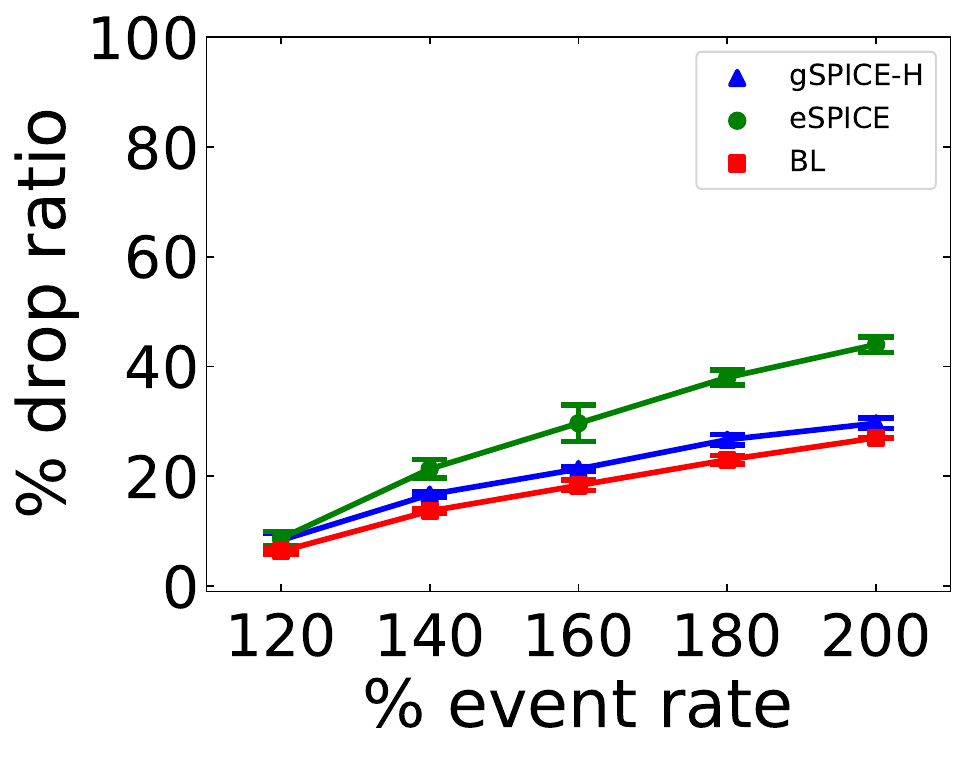}
%		\vspace{-0.6cm}		
		\caption[]{\syntheticQOne: drop ratio}
		\label{fig:synthetic-q1-drop-r}
	\end{subfigure}
%	\vspace{-0.4cm}
	\caption{Synthetic: Impact of event rate on false negatives and drop ratio.}
	\label{fig:synthetic-fn-r}
%	\vspace{-0.4cm}
\end{figure*}

\textbf{Impact on False Negatives.}
Figure \ref{fig:synthetic-fn-r} shows the shedding impact, with different event rates, on the false negatives  for all queries and the ratio of dropped events for query \syntheticQOne.  The drop ratio indicates the overhead of a load shedding strategy. We observed similar results, w.r.t. drop ratio, for \syntheticQTwo{} and \syntheticQFour, hence we do not show them.  
%In the figure, the x-axis represents the event rate. The y-axis in Figures \ref{fig:synthetic-q1-strict-fn-r}, \ref{fig:synthetic-q2-strict-fn-r}, and \ref{fig:synthetic-q4-strict-fn-r} represents the percentage of false negatives while, in Figure \ref{fig:synthetic-q1-drop-r}, it represents the ratio of dropped events.

Increasing the event rate increases the overload on the operator, thus increasing the need to drop more events. Dropping more events might increase the percentage of false negatives. Figure \ref{fig:synthetic-q1-strict-fn-r} depicts results for \syntheticQOne{} showing that the impact of \gSPICESH{} on false negatives is almost negligible irrespective of the used event rate.  In Figure \ref{fig:synthetic-q1-strict-fn-r}, the percentage of false negatives caused by eSPICE and BL increases
% from 19\% to 53\% and 11\% to 45\%
  with increasing the event rates. 

The results show that \gSPICESH{} significantly outperforms, w.r.t. false negatives, eSPICE and BL for \syntheticQOne. That is because \gSPICESH{} uses complex features,  such as type frequency and event attributes, that improve the prediction accuracy. While eSPICE and BL use only simple features such as the event type and position within a window.
Although \gSPICESH{} uses complex features, Figure \ref{fig:synthetic-q1-drop-r} shows that \gSPICESH{} has a relatively low drop ratio compared to other load shedding approaches, where its drop ratio is comparable to the drop ratio of BL. That shows that \gSPICESH{} is a lightweight load shedding approach.

The results for \syntheticQTwo{} are depicted in Figure \ref{fig:synthetic-q2-strict-fn-r}. The percentage of false negatives caused by \gSPICESH{} only slightly increases when increasing the event rate, while those caused by eSPICE and BL increase significantly with increasing event rate.  
Figure \ref{fig:synthetic-q4-strict-fn-r} shows results for query \syntheticQFour. The figure shows that \gSPICESH, again, has a good performance where it results in almost zero false negatives.
Similar to the results of \syntheticQOne, the results show that \gSPICESH{} outperforms, w.r.t. false negatives,  eSPICE and BL for \syntheticQTwo{} and \syntheticQFour.

\begin{figure}[t]
	\centering
	\begin{subfigure}[t]{\figureWidthTwoInRow\linewidth}
		\includegraphics[width=\linewidth]{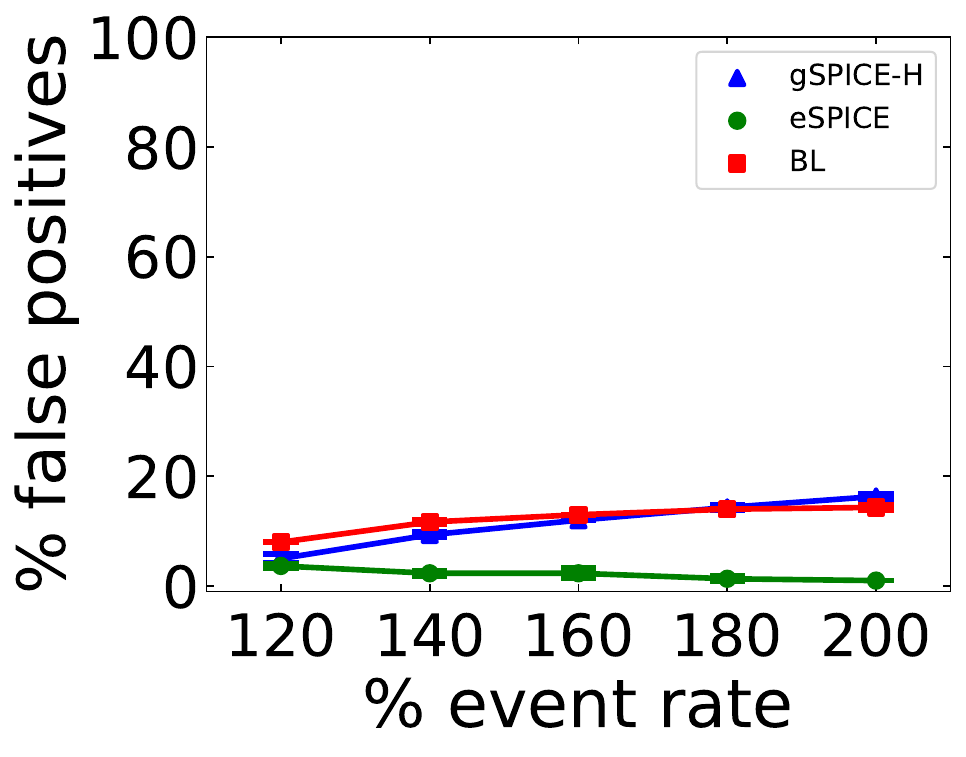} 
%		\vspace{-0.6cm}
		\caption[t]{\syntheticQOne}
		\label{fig:synthetic-q1-strict-fp-r}
	\end{subfigure}
	\hfill%
	\begin{subfigure}[t]{\figureWidthTwoInRow\linewidth}
		\includegraphics[width=\linewidth]{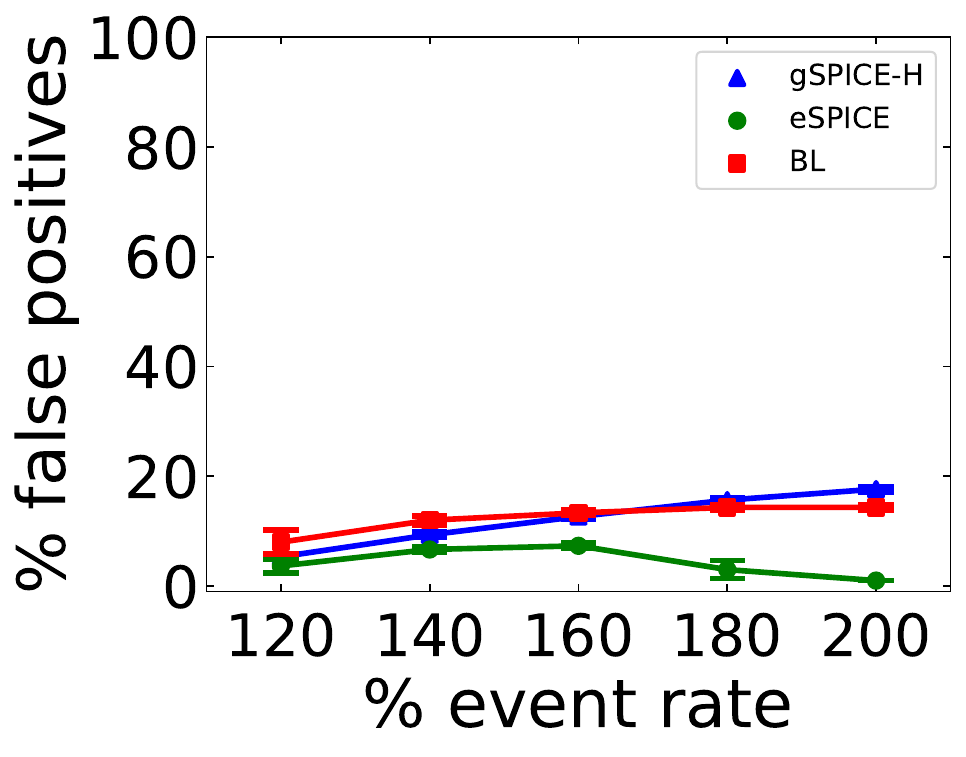} 
%		\vspace{-0.6cm}
		\caption[t]{\syntheticQTwo}
		\label{fig:synthetic-q2-strict-fp-r}
	\end{subfigure}
%	\vspace{-0.4cm}
	\caption{Synthetic: Impact of event rate on false positives.}
	\label{fig:synthetic-fp-r}
%	\vspace{-0.4cm}
\end{figure}

\textbf{Impact on False Positives.}
Figure \ref{fig:synthetic-fp-r} shows the shedding impact  on the false positives for queries \syntheticQOne{} and  \syntheticQTwo.  
%In the figure, the x-axis represents the event rate, while the y-axis represents the percentage of false positives.  
We observe similar results for \syntheticQFour, hence we do not show them. 

Figure \ref{fig:synthetic-q1-strict-fp-r} depicts results for \syntheticQOne{} showing that the percentage of false positives caused by \gSPICESH{} and BL slightly increases when increasing the event rate. 
%Similarly, the percentage of false positives caused  by BL increases with the event rate.
 However, the impact of eSPICE on the percentage of false positives in \syntheticQOne{} is negligible, as shown in Figure \ref{fig:synthetic-q1-strict-fp-r}. eSPICE outperforms, w.r.t. false positives, \gSPICESH{} since eSPICE, in contrast to \gSPICESH, considers the order of events in windows when predicting the event utilities, which has a considerable impact on the false positives. Considering event orders enables eSPICE to assign to event instances of the same event type different utilities depending on their probability to match the pattern.
%  where event instances that are more likely to match the pattern are assigned higher utilities.
Figure \ref{fig:synthetic-q2-strict-fp-r} shows that the results for \syntheticQTwo{} have similar behavior. 

\subsubsection{Stock Results}
\label{sec:stock-results}
Now, we show the results obtained from evaluating \gSPICE{} over the NYSE dataset. We run experiments with queries \stockQOne{} and \stockQThree, where  \gSPICE{} uses a predecessor pane of length 50 events, i.e., $L_\omega= 50$.  

\begin{figure}[t]
	\centering
	\begin{subfigure}[t]{\figureWidthTwoInRow\linewidth}
		\includegraphics[width=\linewidth]{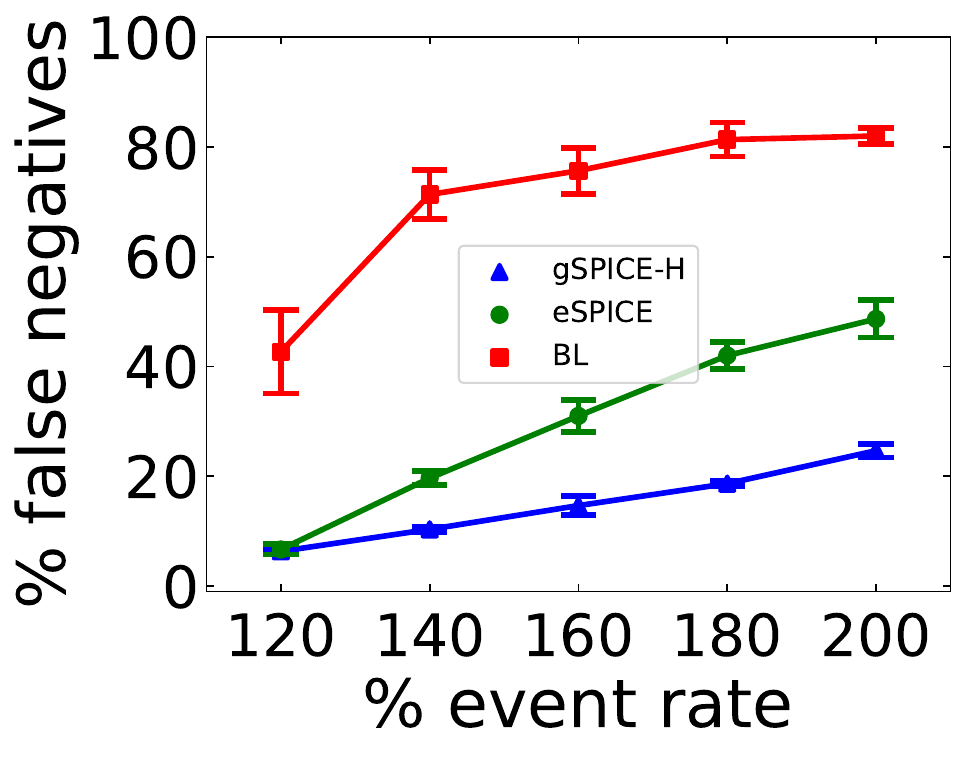}
%		\vspace{-0.6cm} 
		\caption[t]{\stockQOne}
		\label{fig:stock-q1-strict-fn-r}
	\end{subfigure}
	\hfill%
	\begin{subfigure}[t]{\figureWidthTwoInRow\linewidth}
		\includegraphics[width=\linewidth]{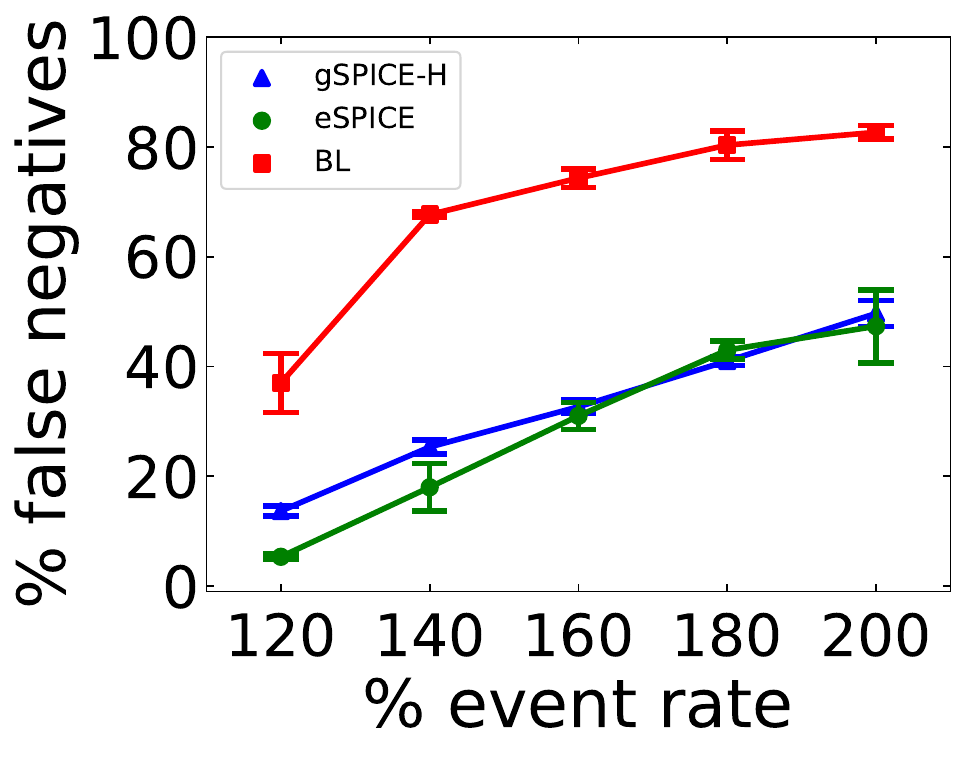}
%		\vspace{-0.6cm}
		\caption[]{\stockQThree}
		\label{fig:stock-q3-strict-fn-r}
	\end{subfigure}
%	\vspace{-0.4cm}
	\caption{Stock: Impact of event rate on false negatives.}
	\label{fig:stock-fn-r}
%	\vspace{-0.6cm}
\end{figure}

\textbf{Impact on False Negatives.}
Figure \ref{fig:stock-fn-r} shows the percentage of false negatives for queries \stockQOne{} and \stockQThree{}. 
%In the figure, the x-axis represents the event rate, while the y-axis represents the percentage of false negatives.
Figure \ref{fig:stock-q1-strict-fn-r} shows results for \stockQOne, where the percentage of false negatives caused by the load shedders increases when increasing the event rate. 
The results show that \gSPICESH{} performs better than BL and eSPICE  by up to 7.2 and 2 times, respectively. Again, the results show that using the type frequency and event attributes in \gSPICESH{} improves the accuracy of predicted event utilities. However, the performance, w.r.t. false negatives, of \gSPICESH{} is worse than its performance when using synthetic data. That is because, in \stockQOne, \gSPICESH{} matches stock events that might have an increase or decrease in their quotes (i.e., attribute values). Hence in \stockQOne, the event attributes provide less useful information to predict the event utilities compared to event attributes in queries on synthetic data.
 
Figure \ref{fig:stock-q3-strict-fn-r} shows results for \stockQThree, where the percentage of false negatives for all load shedders increases when increasing the event rate. The performance of \gSPICESH{} with \stockQThree{} is worse than its performance with \stockQOne{} due to the following.
Since \stockQThree{} contains the negation event operator, \gSPICESH{} might assign, to event types in \stockQThree{} that are before the negated event type (i.e., $C_5$), higher utilities than the event types that are after the negated event type. That might negatively influence the ability of \gSPICESH{} to correctly drop events, thus, increasing its impact on QoR.
Figure \ref{fig:stock-q3-strict-fn-r} shows that  eSPICE outperforms \gSPICESH{} with low input event rates. For an input event rate that is equal to or higher than 160\%, the performance of \gSPICESH{} and eSPICE is comparable.
The results show that \gSPICESH{} has considerably better performance compared to BL.

\begin{figure}[t]
	\centering
	\begin{subfigure}[t]{\figureWidthTwoInRow\linewidth}
		\includegraphics[width=\linewidth]{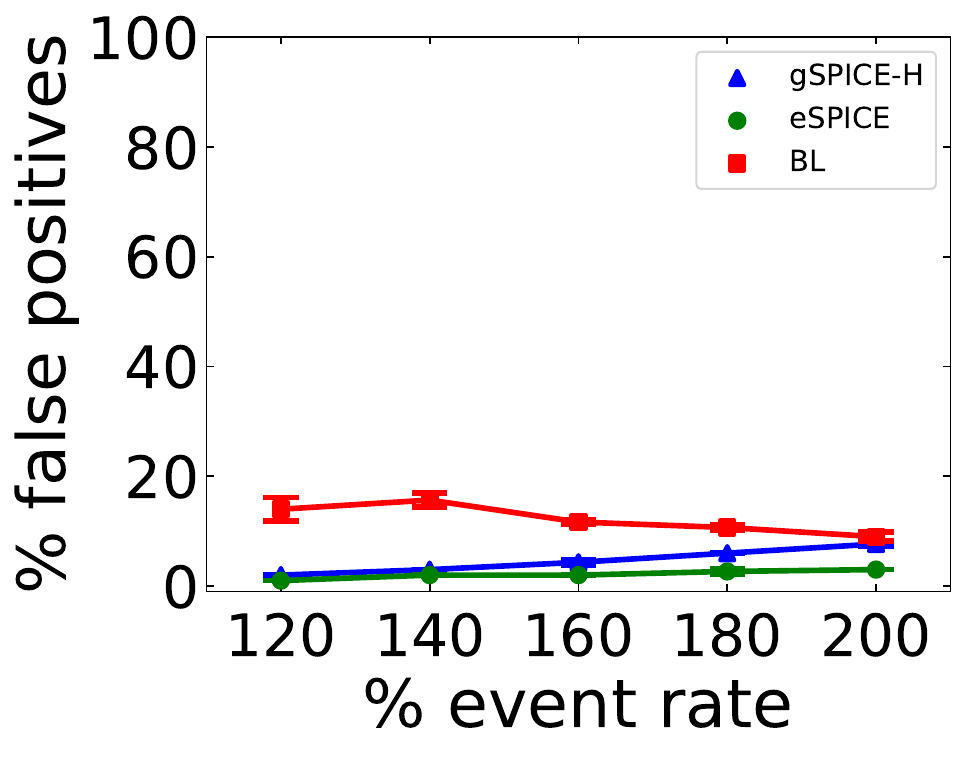} 
%		\vspace{-0.6cm}
		\caption[t]{\stockQOne}
		\label{fig:stock-q1-strict-fp-r}
	\end{subfigure}
	\hfill%
	\begin{subfigure}[t]{\figureWidthTwoInRow\linewidth}
		\includegraphics[width=\linewidth]{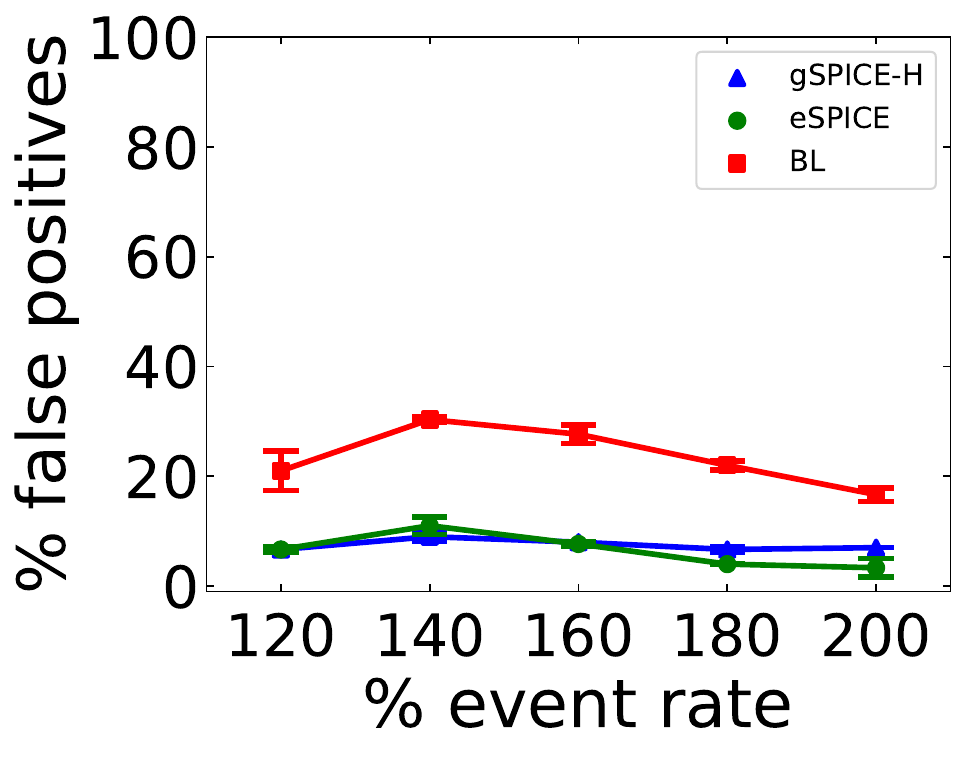}
%		\vspace{-0.6cm} 
		\caption[t]{\stockQThree}
		\label{fig:stock-q3-strict-fp-r}
	\end{subfigure}
%	\vspace{-0.4cm}
	\caption{Stock: Impact of event rate on false positives.}
	\label{fig:stock-fp-r}
%	\vspace{-0.5cm}
\end{figure}

\textbf{Impact on False Positives.}
Figures \ref{fig:stock-q1-strict-fp-r} and \ref{fig:stock-q3-strict-fp-r} show the shedding impact on false positives for queries \stockQOne{} and \stockQThree. 
%In the figure, the x-axis represents the event rate, while the y-axis represents the percentage of false positives.
In Figure \ref{fig:stock-q1-strict-fp-r}, the percentage of false positives caused by \gSPICESH{} and eSPICE slightly increases when the event rate increases. 
Also, the percentage of false positives caused by BL slightly increases when the event rate increases from 120\% to 140\%. After that, the percentage of false positives  decreases when increasing the event rate because with high event rates, BL results in a high number of false negatives, which may imply that only a small number of complex events are detected. That may result in a low percentage of false positives.
%The results for query \stockQTwo{} show similar behavior to the results of \stockQOne, as depicted in Figure \ref{fig:stock-q2-strict-fp-r}. 
For query \stockQThree, results show that the percentage of false positives for all load shedders slightly increases when the event rate increases from 120\% to 140\% (cf. Figure \ref{fig:stock-q3-strict-fp-r}). After that, the percentage of false positives  decreases when increasing the event rate.

\subsubsection{Soccer Results}
\label{sec:soccer-results}
Next, we analyze the performance of \gSPICE{} on the RTLS dataset. Please note, since event types in the RTLS dataset occur periodically (cf. \ref{sec:experimental-setup}), the predecessor pane  $\omega^e$ may not help predict the event utilities as all event types will have, on average, the same frequency in the type frequency $F$.  We run experiments with query \soccerQOne{} using a pane of length 200 events, i.e., $L_\omega= 200$.
%, and  the following event rates: 120\%, 140\%, 160\%, 180\%, and 200\% of the operator throughput $\mu$. 

\begin{figure}[t]
	\centering
	\begin{subfigure}[t]{\figureWidthTwoInRow\linewidth}
		\includegraphics[width=\linewidth]{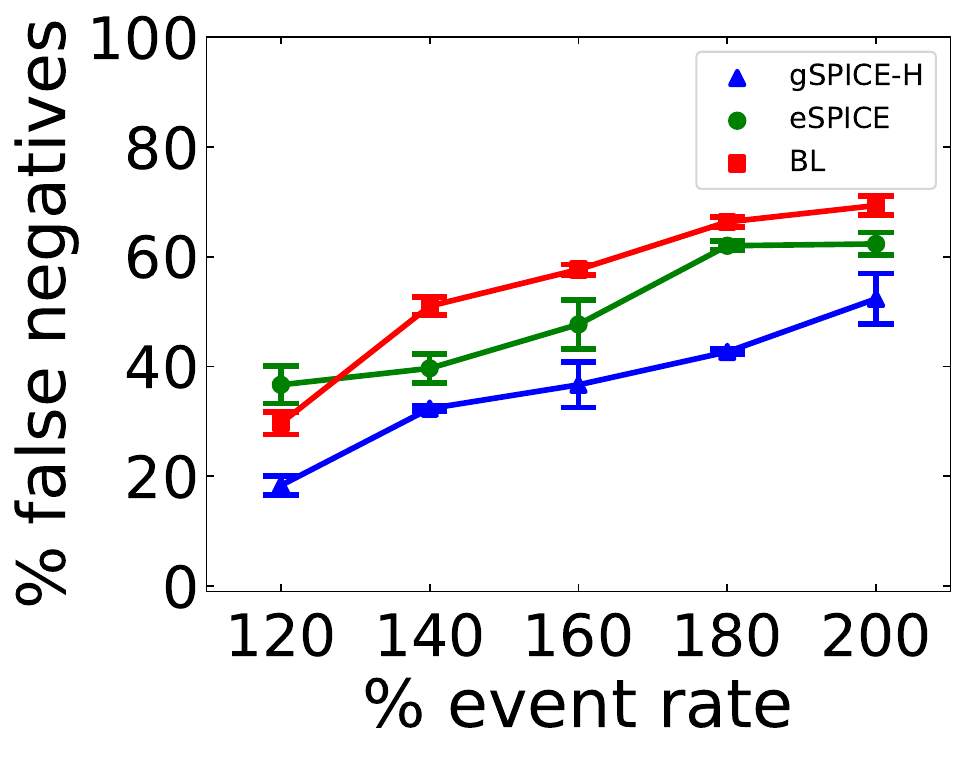}
%		\vspace{-0.6cm} 
		\caption[t]{\soccerQOne: false negatives}
		\label{fig:soccer-q1-strict-fn-r}
	\end{subfigure}
	\hfill%
	\begin{subfigure}[t]{\figureWidthTwoInRow\linewidth}
		\includegraphics[width=\linewidth]{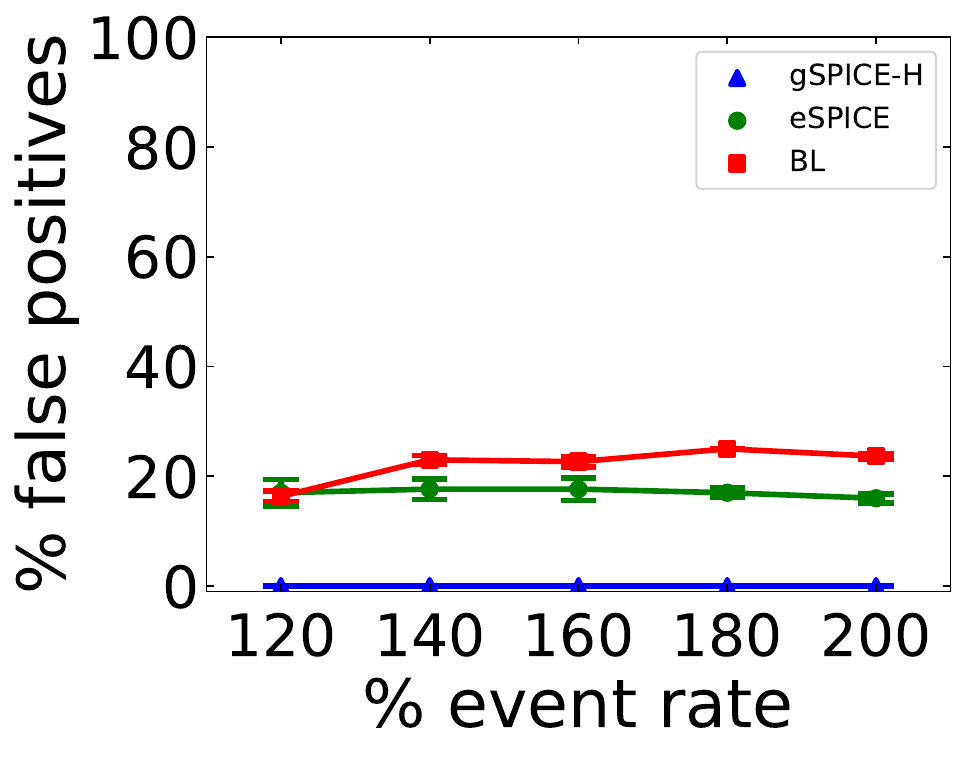}
%		\vspace{-0.6cm} 
		\caption[t]{\soccerQOne: false positives}
		\label{fig:soccer-q1-strict-fp-r}
	\end{subfigure}
%	\vspace{-0.4cm}
	\caption{Soccer: Impact of event rate on QoR.}
	\label{fig:soccer-qor-r}
%	\vspace{-0.4cm}
\end{figure}

Figures \ref{fig:soccer-q1-strict-fn-r} and  \ref{fig:soccer-q1-strict-fp-r} show results for query \soccerQOne.
\soccerQOne{} contains the \textit{any} event operator, where any event type (i.e., any defender from the opposite team) may match the pattern. Hence, the event utilities are more spread out, and it is hard to accurately predict the utilities for different event types.  However, the figure shows that \gSPICESH{} still outperforms BL and eSPICE, irrespective of the used event rate. Moreover, \gSPICESH{} results in almost zero false positives, as depicted in Figure \ref{fig:soccer-q1-strict-fp-r}. The percentage of false positives caused by eSPICE slightly decreases when increasing the event rate, while the impact of BL on false positives slightly increases when increasing the event rate. The results show that \gSPICESH{} has a relatively good performance, w.r.t. QoR, even when the predecessor pane $\omega^e$ is not very useful for predicting the event utilities. That implies that the other two features (i.e., the event type and event attributes) used to predict the event utilities in \gSPICESH{} are important features.

\subsubsection{Impact of Predecessor Pane Length on QoR}
\label{sec:impact-of-context-length}
The pane length may considerably impact the utility prediction. Hence, it may influence the impact of \gSPICE{} on QoR.  For an event $e$, the pane length defines the number of past incoming events that might have an impact on the importance of event $e$. If the length of the predecessor pane is too small, \gSPICE{} may not be able to capture the events that influence the utility of the event $e$. On the other hand, if the length of the predecessor pane is too large, the predecessor pane  $\omega^e$ might contain many unrelated events (i.e., noisy data)  that might hinder accurately predicting the event utilities. Moreover,  a large  predecessor pane might increase the overhead of \gSPICE, thus negatively impacting QoR. 
To evaluate the impact of pane length on the performance, w.r.t. QoR, of \gSPICE{}, we run experiments with \syntheticQTwo, over dataset $DS_5$, and \stockQOne. For \syntheticQTwo, we use a pane  of the following lengths: 5, 10, 20, 40, 80, 320. While for \stockQOne, we use a pane of the following lengths: 10, 50, 100, 400, 800, 1600. Moreover, for both queries,  we use a fixed event rate of 180\% of the operator throughput $\mu$. 

\begin{figure}[t]
	\centering
	\begin{subfigure}[t]{\figureWidthTwoInRow\linewidth}
		\includegraphics[width=\linewidth]{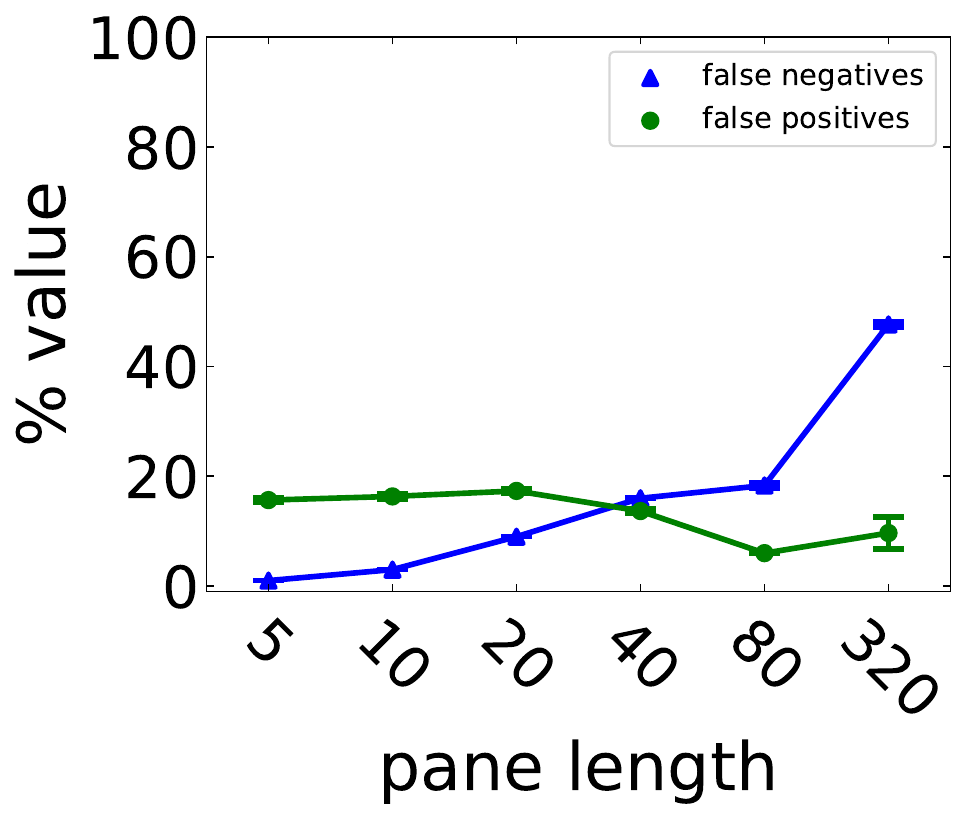}
%		\vspace{-0.6cm}
		\caption[t]{\syntheticQTwo}
		\label{fig:synthetic-q2-strict-fn-fp-c}
	\end{subfigure}
	\hfill%
	\begin{subfigure}[t]{\figureWidthTwoInRow\linewidth}
		\includegraphics[width=\linewidth]{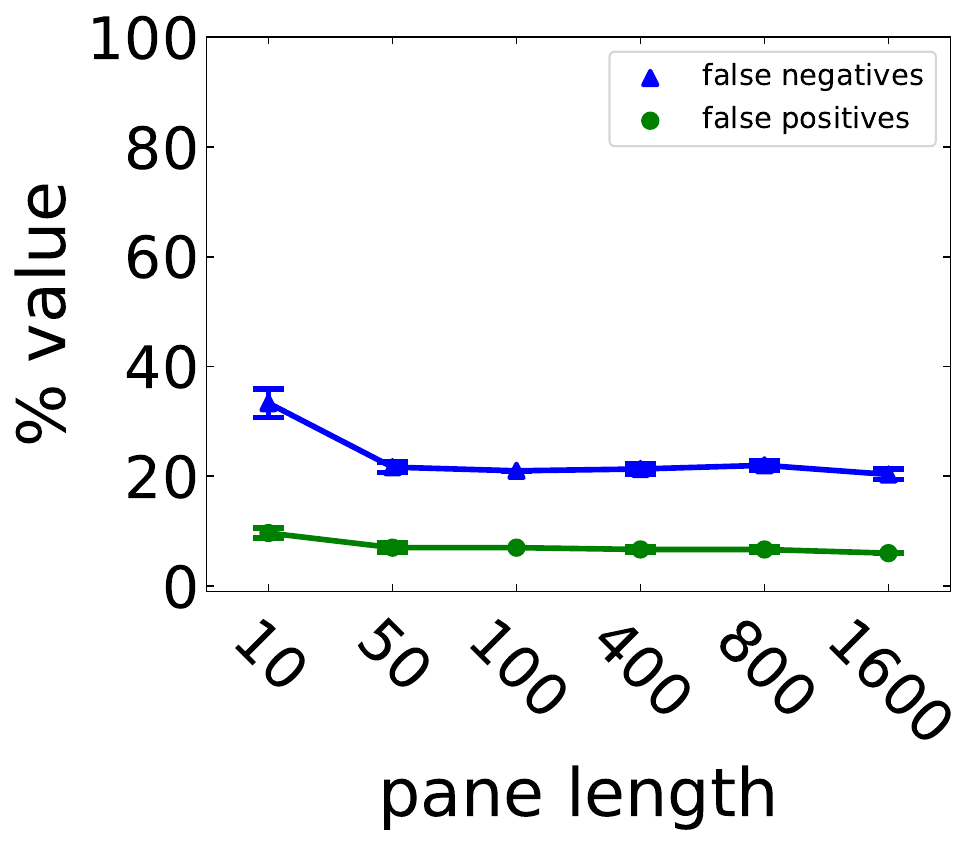}
%		\vspace{-0.6cm}
		\caption[t]{\stockQOne}
		\label{fig:stock-q1-strict-fn-fp-c}
	\end{subfigure}
%	\vspace{-0.4cm}
	\caption{Impact of the predecessor pane length on QoR.}
	\label{fig:qor-c}
%	\vspace{-0.4cm}
\end{figure}

%In the figure, the x-axis represents the predecessor pane length, while the y-axis represents the percentage of false negatives and positives. 
Figure \ref{fig:qor-c} depicts results for both queries.
For \syntheticQTwo, increasing the pane length increases  the percentage of false negatives (cf.  Figure \ref{fig:synthetic-q2-strict-fn-fp-c}).  In the figure, the impact of \gSPICESH{} on the false positives decreases when slightly increasing the pane length. However,  \gSPICESH{} results in more false positives with large pane lengths.  
For  \stockQOne, \gSPICESH{} has a high impact on  the false negatives and positives with small pane lengths (cf. Figure  \ref{fig:stock-q1-strict-fn-fp-c}). However, increasing the pane length thereafter barely changes the incurred false negatives and positives. As a result, we may conclude that using the right pane length may influence the impact of \gSPICESH{} on QoR, where the right pane length depends on the used query and data.

\subsubsection{Impact of Event Distribution on QoR}

The event distribution may considerably impact the performance, w.r.t. QoR, of \gSPICE{} due to the following. The predecessor pane $\omega^e$, which represents an important feature to predict the event utility loses its importance when all event types occur with the same frequency in a dataset. That implies, the type frequency $F$ in the predecessor pane $\omega^e$ will  almost always be the same. Hence, the type frequency will not help predict the event utilities. To evaluate this, we run experiments with all queries on the synthetic data. For all queries, we use a predecessor pane of length 10 events, i.e., $L_\omega= 10$, and a fixed event rate of 140\% of the operator throughput $\mu$.  
As mentioned in Section \ref{sec:predicting_event_utility}, \gSPICE{} may use machine learning models to estimate the event utilities. Therefore, we also show the performance of \gSPICE{} when using a decision tree or a random  forest to predict the event utilities. We refer to \gSPICE{} as \gSPICEST{} and \gSPICESF{} when using a decision tree and a random forest, respectively. 
In our experiments, the random forest consists of ten trees. 
%Figure \ref{fig:synthetic-qor-d} depicts results for query \syntheticQTwo{} and Figure \ref{fig:synthetic-drop-d} shows the corresponding drop ratio.

\begin{figure}[t]
	\centering
	\begin{subfigure}[t]{\figureWidthTwoInRow\linewidth}
		\includegraphics[width=\linewidth]{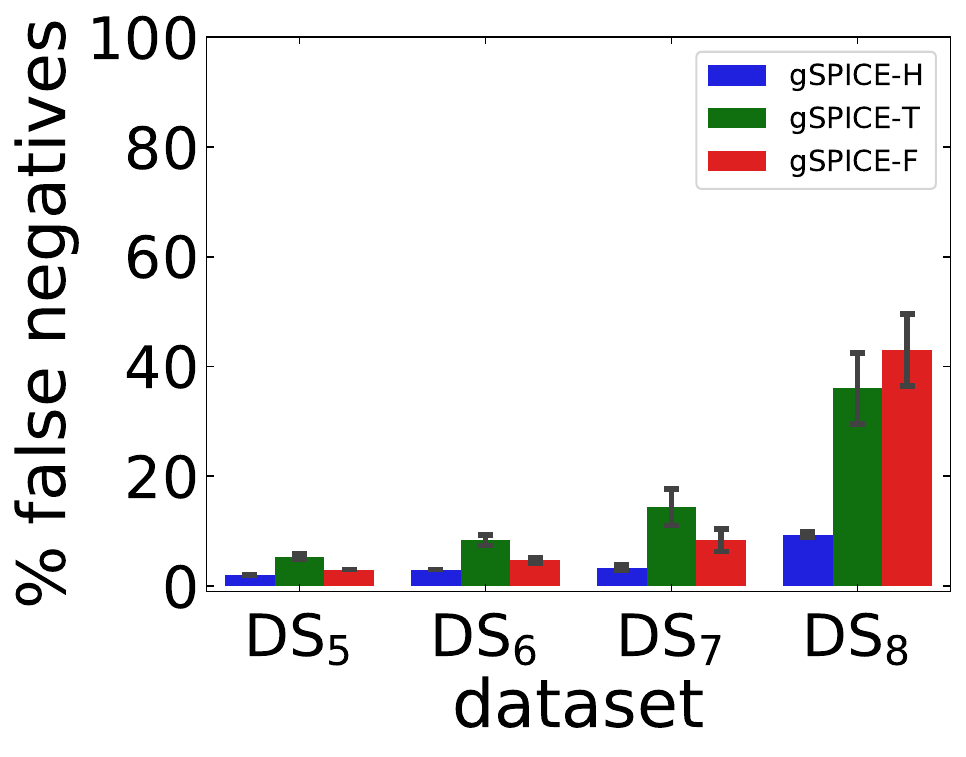} 
%		\vspace{-0.6cm}
		\caption[t]{\syntheticQTwo: false negatives}
		\label{fig:synthetic-q2-strict-fn-d}
	\end{subfigure}
	\hfill%
	\begin{subfigure}[t]{\figureWidthTwoInRow\linewidth}
		\includegraphics[width=\linewidth]{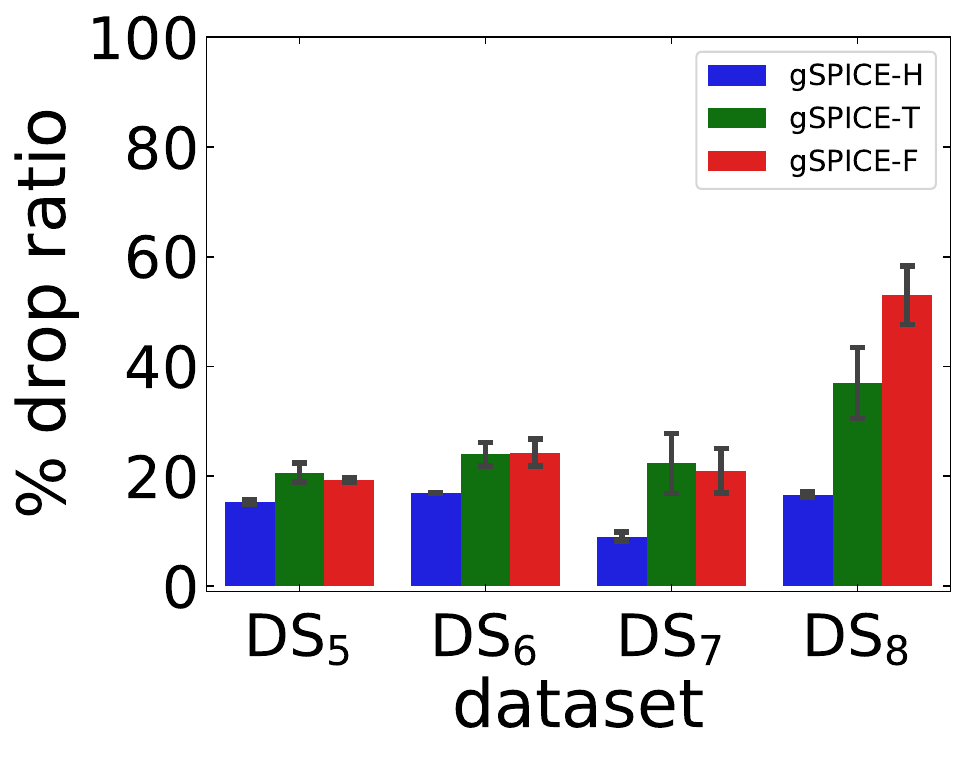} 
%		\vspace{-0.6cm}
		\caption[t]{\syntheticQTwo: drop ratio}
		\label{fig:synthetic-q2-drop-d}
	\end{subfigure}
%	\vspace{-0.4cm}
	\caption{Impact of event distribution.}
	\label{fig:synthetic-qor-d}
%	\vspace{-0.4cm}
\end{figure}

Figure \ref{fig:synthetic-q2-strict-fn-d} depicts false negatives for query \syntheticQTwo, and Figure \ref{fig:synthetic-q2-drop-d} shows the corresponding drop ratio.
We observe similar behavior for false positives and other queries, hence we do not show them.
The results show that for all variants of \gSPICE, the percentage of false negatives is the lowest when using the dataset $DS_5$ and the highest when using the dataset $DS_8$ (cf. Figure \ref{fig:synthetic-q2-strict-fn-d}). In dataset $DS_5$, there exists a high difference between the frequency of event types. For example, events of type $A$ are expected to form 40.7\% of events in $DS_5$,  while events of type $C$ represent only 2.5\% (cf. Table \ref{tab:synthetic-datasets}). This large difference between the amount of each event type enables the predecessor pane $\omega^e$ (i.e., the type frequency $F$) to contain more useful information that helps predict event utilities. While in $DS_8$, all event types occur at the same frequency on average. Hence, for dataset $DS_8$, the predecessor pane  $\omega^e$ is not a useful indicator of the importance of event $e$. Figure \ref{fig:synthetic-q2-strict-fn-d} also shows that using datasets $DS_6$  and $DS_7$, the percentage of false negatives caused by all load shedders, is higher compared to the case when using dataset $DS_5$. 

As Figure \ref{fig:synthetic-q2-strict-fn-d} shows, the performance of \gSPICESF{} is better than the performance of \gSPICEST{} irrespective of the used dataset. Moreover, the performance of \gSPICESF{} and \gSPICESH{} is comparable with datasets $DS_5$ and $DS_6$.
That means that in the case of limited available memory,  \gSPICESF{} might be used as a replacement of \gSPICESH{} with only a slight impact on QoR for these distributions. 
The performance of \gSPICESF{} with $DS_7$ and $DS_8$ is worse than the performance of \gSPICESH, especially with $DS_8$. Moreover, \gSPICESH{} outperforms, w.r.t. false negatives,  \gSPICEST, irrespective of the used dataset.
That is because \gSPICESF{} and \gSPICEST{} result in a high drop ratio compared to \gSPICESH{} (cf. Figure \ref{fig:synthetic-q2-drop-d}). 
%A high drop ratio implies that more events are dropped, hence negatively impacting QoR. 

\subsubsection{Maintaining Latency Bound}
\begin{figure}[t]
	\centering
	\begin{subfigure}[t]{\figureWidthTwoInRow\linewidth}
		\includegraphics[width=\linewidth]{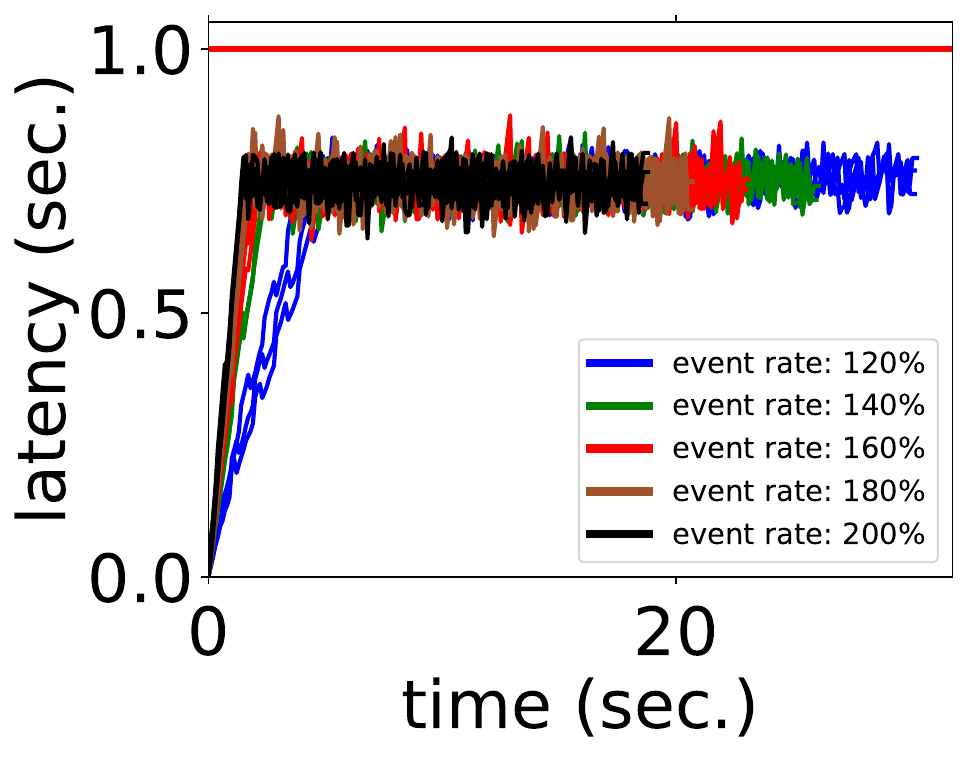}
%		\vspace{-0.6cm} 
		\caption[t]{\syntheticQOne}
		\label{fig:q1-latency-r}
	\end{subfigure}
	\hfill%
	\begin{subfigure}[t]{\figureWidthTwoInRow\linewidth}
		\includegraphics[width=\linewidth]{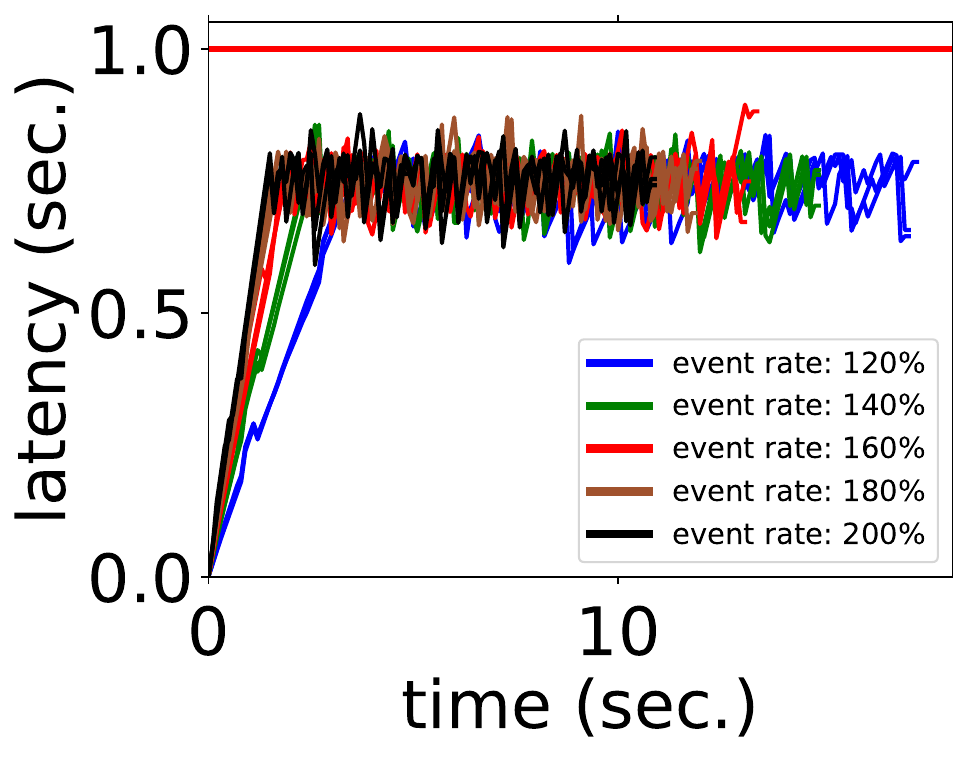}
%		\vspace{-0.6cm}
		\caption[]{\stockQOne}
		\label{fig:q6-latency-r}
	\end{subfigure}
%	\vspace{-0.2cm}
	\caption{Maintaining latency bound.}
	\label{fig:latency-r}
%		\vspace{-0.4cm}
\end{figure}

\gSPICE{} performs load shedding to maintain a given latency bound (LB). Figure \ref{fig:latency-r} shows the ability of \gSPICE{} to maintain the given latency bound, where it depicts results for \syntheticQOne{} and \stockQOne. We observe similar results for other queries, hence we do not show them. For all queries, we use the same setting as explained in Sections \ref{sec:impact-of-event-rate}, \ref{sec:stock-results}, and \ref{sec:soccer-results}. 
%In Figure \ref{fig:latency-r}, the x-axis represents the elapsed time, and the y-axis represents the induced event latency.
The figure shows that \gSPICE{} always maintains the given latency bound, irrespective of the event rate. The induced event latency stays around 800 milliseconds (i.e., 80\% of LB that represents a safety bound). This shows that \gSPICE{} can successfully maintain a given latency bound.

\subsubsection{Discussion}
Through extensive evaluations with several datasets and representative queries, we see that for the majority of queries and datasets, \gSPICE{} outperforms  state-of-the-art black-box load shedding approaches w.r.t. QoR.  \gSPICE{} performs especially well when the event types do not follow a uniform distribution and when using the sequence event operator.
However, for low input event rates, eSPICE outperforms (w.r.t. QoR) \gSPICE{} when using the negation event operator.
Moreover, the results show that using the right predecessor pane length may considerably improve the performance of \gSPICE{} . Further, the overhead of performing load shedding in \gSPICE{} is relatively low, hence \gSPICE{} is  lightweight. 
Also, the results show that to reduce the required memory, \gSPICE{} may use well-known machine learning models, e.g.,  random forests, with a slight adverse impact on QoR.

\section{Related Work}

 Load shedding has been extensively studied in the stream processing domain \cite{Carney:2002:MSN:1287369.1287389, Tatbul:2003:LSD:1315451.1315479, Olston:2003:AFC:872757.872825, Ayad:2004:SOC:1007568.1007616, Wei:2010:AHO:1920841.1920998, motwani2003query, 8622265}. 
The idea is to drop tuples to reduce the system load but still provide the maximum possible QoR. Hence, the crucial question here is which tuples to drop so that QoR is not impacted drastically. In \cite{Carney:2002:MSN:1287369.1287389, Tatbul:2003:LSD:1315451.1315479, 8622265}, the authors assumed that  tuples have different utilities and impact on QoR where the utility of tuples depends on their content.  In case of overload, tuples with low utility values are dropped. In \cite{Carney:2002:MSN:1287369.1287389, Tatbul:2003:LSD:1315451.1315479},  the authors assume that the mapping between the utility and tuple's content is given, for example, by an application expert, while, in \cite{8622265, Olston:2003:AFC:872757.872825}, they learn this mapping online depending on the used query.

The work in \cite {Gedik:2005:ALS:1099554.1099587, 4221702} propose load shedding approaches for join operators where the goal is to increase the number of output  tuples.  In all the above works, the utilities of tuples  are either computed using simple dependencies between tuples (i.e., in join operators) or they are computed  for each tuple individually without considering the dependency between tuples at all. However, in CEP systems, patterns are more complex than a simple binary join where a pattern can be viewed as multi-relational non-equi-joins  with temporal constraints. Moreover, events in a pattern are interdependent with each other which we must take into consideration when assigning utilities to events.  

There exist several works on load shedding in CEP \cite{He2014OnLS, espice, pspice, bo:2020}. The main goal  of these approaches is to drop events or PMs in overload cases to prevent violating a given latency bound or crashing the system. In \cite{He2014OnLS, espice}, the authors propose black-box load shedding approaches that drop events in overload cases. While in \cite{pspice, bo:2020}, the authors propose white-box load shedding approaches.   In overhead cases, the approaches in \cite{pspice, bo:2020} drop PMs. Moreover, in \cite{bo:2020}, the authors propose to drop events as well. 
%In \cite{hspice-tbd}, the authors propose a load shedding approach that drops events either from windows or from PMs.
 All these approaches depend on features, such as event type, event position in the window, and PM progress to predict the utility of events and PMs.  The work in \cite{bo:2020} also uses  event attributes to predict the utility of events. 
%  They assign a utility value to an event depending on the PMs to which the event contributes. 
  If the event contributes to a low importance PM, the event is assigned a low utility. However, an event may, simultaneously, contribute to low and high importance PMs which the authors do not consider. Moreover,  in \gSPICE, our proposed utility prediction model uses event attributes very differently.
Also, in contrast to all these works, \gSPICE{} uses the predecessor pane of an event as a feature to predict the event utilities, where the importance of an event might heavily depend on the prior occurred events. Moreover, we show how to  efficiently use event attributes as a feature to predict the event utility in \gSPICE{}.

\section{Conclusion}
\label{sec:conclusion}
In this paper, we proposed an efficient black-box load shedding approach, called \gSPICE{}, that drops events from the input event stream to maintain a given latency bound in the presence of overload. 
To predict event utilities, \gSPICE{} uses a probabilistic model  that depends on the following features: 1) event type, 2) type frequency in the predecessor pane, and 3) event attributes. \gSPICE{} uses Zobrist hashing to efficiently store event utilities. Moreover, if the utility table is very large, to minimize the needed memory for storing utilities, \gSPICE{}  uses a machine learning model (e.g., decision trees or random forests) to estimate event utilities.
Through extensive evaluations on  several representative queries and several real-world and synthetic datasets, we show that, for the majority of cases, \gSPICE{} outperforms state-of-the-art black-box load shedding approaches, w.r.t. QoR.
Moreover, the results prove that \gSPICE{} is a lightweight shedding approach.
 Additionally, we show that \gSPICE{} always maintains the given latency bound regardless of the incoming input event rate.

\balance
\section*{Acknowledgement}
This work was supported by the German Research Foundation (DFG) under the research grant "PRECEPT II" (BH 154/1-2 and RO 1086/19-2).

\bibliographystyle{IEEEtran}
\bibliography{paper}

\end{document}